\documentclass[aps,amsmath,amssymb,nofootinbib]{revtex4} 
\usepackage{graphicx} 
\usepackage{amsmath}
\usepackage{amsfonts,amsbsy}
\usepackage{amssymb}
\usepackage{relsize}
\allowdisplaybreaks

\usepackage{xcolor}
\definecolor{lcolor}{rgb}{0.5,0,0}
\definecolor{citcolor}{rgb}{0,0.3,0.0}
\usepackage[breaklinks,colorlinks,urlcolor=blue,citecolor=citcolor,linkcolor=lcolor]{hyperref}

\usepackage{mciteplus}

\def\lsim{ \,\, \vcenter{\hbox{$\buildrel{\displaystyle <}\over\sim$}}
 \,\,}
\def\be{\begin{equation}}
\def\ee{\end{equation}}
\def\bea{\begin{eqnarray}}
\def\eea{\end{eqnarray}}

\newcommand{\dd}{{\rm d}}
\newcommand{\nn}{\nonumber}
\newcommand{\tr}{\, \mathrm{tr} }

\begin{document}

\title{Quark and gluon entanglement in the proton on the light cone
at intermediate $x$}

\author{Adrian Dumitru}
\email{adrian.dumitru@baruch.cuny.edu}
\affiliation{Department of Natural Sciences, Baruch College, CUNY,
17 Lexington Avenue, New York, NY 10010, USA}
\affiliation{The Graduate School and University Center, The City University
  of New York, 365 Fifth Avenue, New York, NY 10016, USA}

\author{Eric Kolbusz}
\email{ekolbusz@gradcenter.cuny.edu}
\affiliation{Department of Natural Sciences, Baruch College, CUNY,
17 Lexington Avenue, New York, NY 10010, USA}
\affiliation{The Graduate School and University Center, The City University
  of New York, 365 Fifth Avenue, New York, NY 10016, USA}

\begin{abstract}
In QCD with $N_c$ colors the anti-symmetric valence quark color space
singlet state $\sim \epsilon_{i_1\cdots i_{N_c}} |i_1,\cdots,
i_{N_c}\rangle$ of the proton corresponds to the reduced density
matrix $\rho_{ij}=(1/N_c) \delta_{ij}$ for a single color degree of
freedom. Its degenerate spectrum of eigenvalues, $\lambda_i=1/N_c$, the
purity $\tr \rho^2 = 1/N_c$, and the von~Neumann entropy
$S_\mathrm{vN}=\log(N_c)$ all indicate maximal entanglement of color.

On the other hand, for $N_c\to\infty$ the spatial wave function of the proton
factorizes into valence quark wave functions determined by a mean
field (E.~Witten, Nucl.\ Phys.\ B 160 (1979) p.\ 57) where there is no
entanglement of spatial degrees of freedom.

A model calculation at $N_c=3$ using a simple three quark
light-front wave function by Brodsky and Schlumpf, predicts percent
level entanglement of spatial degrees of freedom.

Using light-cone perturbation theory we also derive the density matrix
associated with the four parton $|qqqg\rangle$ Fock state.  Tracing
out the quarks, we construct the reduced density matrix for the
degrees of freedom of the gluon, which encodes its entanglement with
the sources.  Our expressions provide the dependence of the density
matrix on the soft cutoff $x$ for the gluon light-cone momentum,
and on the collinear and ultraviolet regulators.  Numerical
results obtained in a simple approximation indicate stronger
entanglement for the gluon (with $x_g < \langle x_q\rangle$) than for
quarks in the three quark Fock state.
\end{abstract}

\maketitle
\tableofcontents
\newpage

\section{Introduction}

The near future may present exciting opportunities to search
experimentally for color entanglement in QCD~\cite{Lecture-Aidala}.
For example, color entanglement has been argued to break Transverse
Momentum Dependent QCD factorization in the production of hadrons with
high transverse momentum (and a transverse momentum imbalance) in
proton-proton
collisions~\cite{Mulders:2011zt,Rogers:2010dm,Aidala:2021pvc}.  It has
also been proposed that colored quarks and gluons in the wave function
of the proton are
entangled~\cite{Kharzeev:2021nzh,Kharzeev:2017qzs,Kovner:2015hga},
these references focusing specifically on gluons with small light-cone
momentum fractions $x$, and that the entropy in the final state of
high-energy Deeply Inelastic electron-proton Scattering (DIS)
experiments, or of hadronic collisions, may reflect their entanglement
(entropy)~\cite{Kharzeev:2017qzs,Kovner:2018rbf}. This proposal is
currently under active
investigation~\cite{Tu:2019ouv,Kharzeev:2021yyf,Ramos:2020kaj,Hentschinski:2021aux,Zhang:2021hra,H1:2020zpd}.
A non-zero ``entropy of ignorance'' could also arise without tracing
over entangled degrees of freedom, however, just from the fact that
the available measurements provide only limited information on the
density matrix, effectively ``zeroing out'' some of its matrix
elements~\cite{Duan:2020jkz}.  \\

The density matrix and the entanglement entropy of small-$x$ gluons in
a hadron has been computed in the ``Color Glass Condensate'' (CGC)
framework for high-energy QCD in ref.~\cite{Kovner:2015hga}; see,
also, ref.~\cite{Hagiwara:2017uaz} for a relation of the von~Neumann
entropy of small-$x$ gluons to their quantum phase space (Wigner)
distribution. The computation of Kovner and
Lublinsky~\cite{Kovner:2015hga} has been generalized in
ref.~\cite{Duan:2021clk} to the ``high density saturation regime'' at
low transverse momentum (where quasi-particles emerge).

The evolution of the density matrix for soft small-$x$ gluons with
rapidity $Y= \log 1/x$ has been derived recently in
ref.~\cite{Armesto:2019mna}. This is obtained by tracing over gluons
with rapidity less than $Y$ (or light-cone momentum fractions greater
than $x$). The authors find that this evolution equation is of
Lindblad form, describing the non-unitary evolution of the density
matrix of an open system. The purity of the density matrix decreases
with increasing rapidity $Y$.

Ref.~\cite{Dvali:2021ooc} describes a correspondence at weak coupling
between highly occupied black hole states of soft gravitons and the
state of high gluon occupation numbers encountered in the proton at
small $x$. Dvali and Venugopalan argue that upon tracing out the
sources at higher $x$ the entropy of soft degrees of freedom attains
its maximal value permitted by unitarity, and that it is proportional
to the area times a Goldstone scale squared.\\

A substantial amount of work has been done to understand the regime of
high gluon occupation number at small light-cone momentum fraction
$x$, as we have just outlined\footnote{We refer to
  ref.~\cite{Morreale:2021pnn} for a recent review of collider
  searches for non-linear gluon dynamics.}.  Our approach here is
complementary in that we consider the regime of relatively large $x$
where the proton may be composed of only a few particles
(``partons''). In sec.~\ref{sec:DM_O-g2} we consider the emission of
one single gluon from a three-quark leading Fock state.  We will not
require the gluon to be soft and so we recover the dependence of the
density matrix on its light-cone momentum fraction. We shall also
obtain the dependence of the density matrix on the collinear
regulator, which in ref.~\cite{Kovner:2015hga} is implicit in their
parameter $\mu^2$ (the average color charge density squared per unit
transverse area, see
refs.~\cite{Dumitru:2020gla,Dumitru:2021tvw,Dumitru:2021tqp}). \\

Before considering gluon emission, however, in the next
sec.~\ref{sec:qqqDM-LO} we first approximate the proton (for $N_c=3$
colors) by a three-quark state. The color-space wave function $\sim
\epsilon_{n_1 n_2 n_3}$ corresponds to maximal entanglement of color,
in that the reduced density matrix $\rho_{nn'}$ obtained after tracing
out all other degrees of freedom has a degenerate spectrum of
eigenvalues $\lambda_1 = \lambda_2 = \lambda_3 = \frac{1}{3}$. We then
compute reduced density matrices over various spatial (momentum)
degrees of freedom. Using a three-quark light-cone wave function from
the literature~\cite{Schlumpf:1992vq,Brodsky:1994fz}, we find
numerically that these density matrices exhibit a high purity $\tr\,
\rho^2 > 0.95$ and low von~Neumann entropy $S_\mathrm{vN} \lsim 0.15$
(nats).  In this regard, these model wave functions are close to the
$N_c\to\infty$ limit (at fixed $g^2 N_c$) where the spatial wave
function of the proton factorizes into valence quark wave
functions~\cite{Witten:1979kh}.  \\

Kharzeev has argued~\cite{Kharzeev:2021nzh} that the scattering of a
probe located at $x^- =0$ off the proton would lead to ``information
scrambling'' and suppression of off-diagonal elements of the density
matrix of the proton due an average over the phase of its wave
function (see, also, sec.~III in ref.~\cite{Kovner:2018rbf}). The
issue of entanglement and entropy production in {\em particle
  production in high-energy collisions} has been addressed also in
refs.~\cite{Rogers:2010dm,Kovner:2015hga}, and has since been
revisited in some of the references mentioned above. Here, however, we
consider entanglement of various degrees of freedom in a proton {\it
  per se}.  \\

Finally, let us mention that $n$-body quantum correlations manifest
also in a non-trivial impact parameter and transverse momentum
dependence of color charge correlations in the
proton~\cite{Dumitru:2018vpr,Dumitru:2020fdh,Dumitru:2021tvw,Dumitru:2021tqp},
as well as in Bose-Einstein correlations of small-$x$
gluons~\cite{Altinoluk:2015uaa,Kovner:2021lty}. The focus of the
current paper is on entanglement of degrees of freedom in the
proton.\\

The remainder of the paper is organized as follows. In the following
sec.~\ref{sec:qqqDM-LO} we consider the density matrix of the
three-quark Fock state. In sec.~\ref{sec:DM_O-g2} we compute
the leading perturbative correction and the density matrix for the
Fock state containing three quarks and a gluon. Sec.~\ref{sec:summary}
contains a brief summary.

\section{Density Matrix for the three quark Fock state}
\label{sec:qqqDM-LO}

In the absence of gluons protons are made of $N_c=3$ ``valence''
quarks and we can write the proton state $|P\rangle = |P^+,\vec P=0\rangle$
as\footnote{For an introduction into the light-front formalism and its
  application in QCD we refer to
  refs.~\cite{Lepage:1980fj,Harindranath:1996hq,Brodsky:1997de,Brodsky:2000ii,Burkardt:1995ct}.
  We write three momenta as $k=(k^+,\vec k) = (xP^+,\vec k)$ where $x$
  corresponds to the fractional light cone momentum and $\vec k$ to the
transverse momentum.}
\be \label{Pstate_short}
|P\rangle = \int\limits_{[0,1]^3} [\dd x_i] \int [\dd^2 k_i]\,
\Psi_\mathrm{qqq}\left(k_1; k_2; k_3\right)\,\,
\left|k_1; k_2; k_3\right>~,
\ee
where
\bea \label{eq:dxi_d2ki}
\left[\dd x_i\right] &=& \delta\left(1-\sum_i x_i\right) \,\,
  \prod_{i}\frac{\dd x_i}{2x_i}~,\\
  \left[\dd^2 k_i\right] &=& 4\cdot
  16\pi^3\, \delta\left(\sum_i \vec k_i\right) \,\,
  \prod_{i}\frac{\dd^2 k_i}{16\pi^3}~.
\eea
$k_1, k_2, k_3$ denote the ``coordinates'' (light-cone momentum
fractions and transverse momenta) for the three quarks.  Above we only
write the (symmetric\footnote{That is, $\Psi_\mathrm{qqq}$ is
  symmetric under exchange of any two quarks:
  $\Psi_\mathrm{qqq}\left(k_1; k_2; k_3\right) =
  \Psi_\mathrm{qqq}\left(k_2; k_1; k_3\right)$ etc.}) spatial wave
function $\Psi_\mathrm{qqq}\left(k_1; k_2; k_3\right)$ which is our
focus in this section. The color, flavor, spin wave function is
discussed below in sec.~\ref{sec:color-flavor-spin}. Restricting to the
three quark state corresponds to a light front constituent quark model.

We normalize the proton state as
\be
\left< K \, | \, P \right> = 16\pi^3 \, P^+\, \delta(P^+ - K^+)
\, \delta(\vec P - \vec K)~.
\ee
With the standard normalization of quark states,
\be
\langle p\, |\, k\rangle = 16\pi^3\, k^+\,
\delta(p^+ - k^+)\, \delta(\vec p -\vec k)
\ee
this leads to the following normalization condition for the
three-quark wave function:
\be \label{eq:Norm_psi3}
\frac{1}{2} \int [\dd x_i] \int [\dd^2 k_i]\,\,\,
  |\Psi_\mathrm{qqq}|^2 = 1~.
\ee
\\

In what follows it will be useful to factor out the center of momentum
(COM) constraint by transforming to the new
coordinates~\cite{Bakker:1979eg}
\bea
\xi = \frac{x_1}{x_1+x_2}~~,~~ \eta=1-x_3=x_1+x_2~,\nn\\
\vec Q = - \vec k_3~~,~~
\vec q = \frac{1}{2}(\vec k_1 - \vec k_2 + \vec k_3)
+ \vec Q (1-\xi)~,
\eea
with $0 < \xi,\eta < 1$. Then
\be
2\left[\dd x_i\right] = \frac{\dd\xi}{2\xi(1-\xi)}
\, \frac{\dd\eta}{2\eta(1-\eta)}~~~,~~~
\frac{1}{4}\left[\dd^2 k_i\right]
= \frac{\dd^2 q}{16\pi^3}\, \frac{\dd^2 Q}{16\pi^3}~,
\ee
and
\be \label{eq:Norm_Psi}
\int \frac{\dd\xi}{2\xi(1-\xi)}
\, \frac{\dd\eta}{2\eta(1-\eta)}\,\frac{\dd^2 q}{16\pi^3}\,
\frac{\dd^2 Q}{16\pi^3}    \,\,
  |\Psi_\mathrm{qqq}|^2 = 1~.
\ee
\\

\begin{figure}[htb]
  \includegraphics[width=0.44\textwidth]{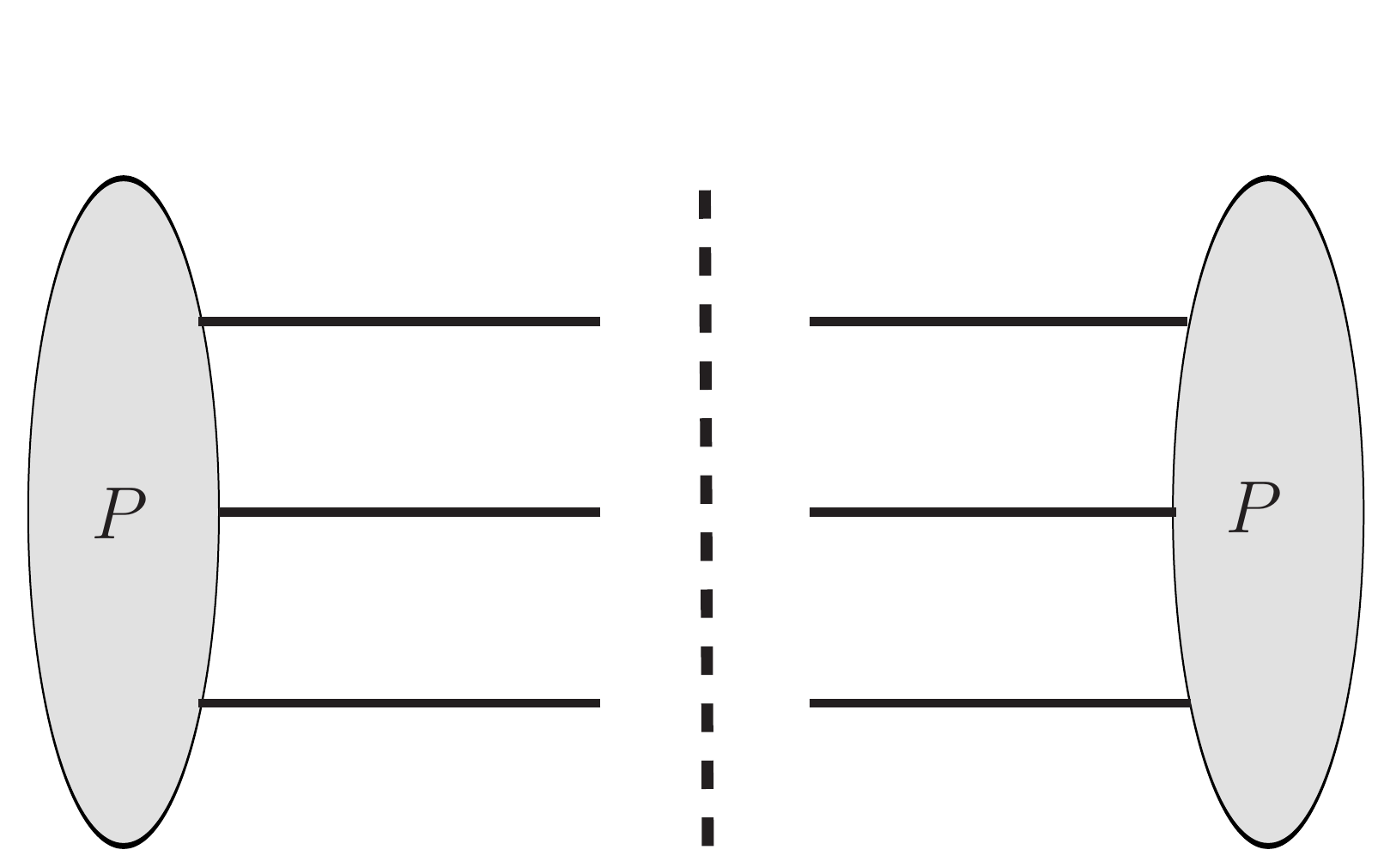}
  \caption{\label{fig:qDM-diagram} Diagrammatic illustration of the
    density matrix $\rho_{\alpha \alpha'}$ for the three-quark
    state. The dashed vertical line indicates the insertion of
    $|\alpha'\rangle\langle\alpha|$ into $\langle P|P\rangle$, where
    $\alpha, \alpha'$ denote sets of longitudinal and transverse
    momenta, and colors of the quarks in $|P\rangle$ and $\langle P|$,
    respectively.  }
\end{figure}
The density operator is $\hat\rho = |P\rangle\, \langle P|$.
To obtain its matrix elements we project the proton state $|P\rangle$
and its dual $\langle P|$ on three-quark states $\alpha\equiv
\left\{x_i,\vec k_i\right\} = \left\{\xi,\eta,\vec q, \vec Q\right\}$
and $\alpha'\equiv \left\{x_i',\vec k_i'\right\} =
\left\{\xi',\eta',\vec q', \vec Q'\right\}$, respectively:
\bea
\rho_{\alpha \alpha'} ~ (2\pi)^3 \,\delta\left(1-\sum_i x_i\right)\,\,
\delta\left(\sum_i \vec k_i\right)~
(2\pi)^3 \, \delta\left(1-\sum_i x_i'\right)\,\,
\delta\left(\sum_i \vec k_i'\right)
= \langle\alpha|P\rangle~ \langle P|\alpha'\rangle~.
\label{eq:Def-rho_pure}
\eea
This is shown diagrammatically in fig.~\ref{fig:qDM-diagram}.
With\footnote{In eq.~(\ref{eq:Def-rho_pure}) we factor out the product
  of $\delta$-functions for the COM constraints so that the density
  matrix satisfies the familiar normalization condition $\tr_\alpha\,
  \rho=1$, eq.~(\ref{eq:tr_alpha-rho_qqq}).}
\bea
\langle\alpha|P\rangle = \Psi_\mathrm{qqq}(\alpha) \,\,
(2\pi)^3 \,\delta\left(1-\sum_i x_i\right)\,\,
\delta\left(\sum_i \vec k_i\right)~,
\label{eq:<alpha|P>}
\eea
finally, the density matrix describing the pure (three-quark) state
is given by
\bea \label{eq:rho-pure}
\rho_{\alpha \alpha'} = \Psi^*_\mathrm{qqq}(\alpha')~ \Psi_\mathrm{qqq}(\alpha) ~.
\eea
Its trace over all degrees of freedom is equal to
\be  \label{eq:tr_alpha-rho_qqq}
\tr_\alpha \, \rho_{\alpha\alpha} \equiv \tr_{\xi,\eta,\vec q,\vec Q}\,
\rho_{\alpha\alpha} \equiv \int \frac{\dd\xi}{2\xi(1-\xi)} \,
\frac{\dd\eta}{2\eta(1-\eta)}\, \frac{\dd^2 q}{16\pi^3}\, \frac{\dd^2
  Q}{16\pi^3} ~\left| \Psi_\mathrm{qqq}(\xi,\eta,\vec q, \vec Q)\right|^2 = 1~,
\ee
where
\be  \label{eq:dalpha_qqq}
\dd\alpha\equiv \frac{\dd\xi}{2\xi(1-\xi)} \,
\frac{\dd\eta}{2\eta(1-\eta)}\, \frac{\dd^2 q}{16\pi^3}\, \frac{\dd^2
  Q}{16\pi^3}
= \frac{1}{2}\, [\dd x_i]\, [\dd^2 k_i]~.
\ee
To arrive at eq.~(\ref{eq:tr_alpha-rho_qqq}) we define the trace of the
density operator via
\bea \label{eq:tr_rho^}
\tr \hat \rho &=& \frac{\int\dd\alpha \int\dd\alpha'\,\, \rho_{\alpha\alpha'}
  \, \langle\alpha' | \alpha\rangle}{\frac{1}{4}\, \langle P|P\rangle} \\
&=&
\frac{\frac{1}{2}\int[\dd x_i]\, [\dd^2 k_i] \,\,
  \frac{1}{2}\int[\dd x_i']\, [\dd^2 k_i'] \,\, \rho_{\alpha\alpha'}
  \, \langle x_1',\vec k_1';\, x_2',\vec k_2';\, x_3',\vec k_3' |
  x_1,\vec k_1; \,x_2,\vec k_2;\, x_3,\vec k_3\rangle
}{\frac{1}{4}\, \langle P|P\rangle}
~,
\eea
and use
\bea
\langle\alpha' | \alpha\rangle = 
(16\pi^3)^3 \, k_1^+\,k_2^+\,k_3^+\,\delta(k_1'-k_1)
\,\delta(k_2'-k_2)\,\delta(k_3'-k_3)~, \nn\\
\int\dd\alpha'\, \langle\alpha' | \alpha\rangle
= \frac{1}{2} \, (2\pi)^3\, \delta(1-\sum x_i)\, \delta(\sum \vec k_i)~.
\label{eq:<alpha'|alpha>-qqq}
\eea
\\

Of course, the above pure density matrix is idempotent:
\be
\int \dd\alpha' \, \rho_{\alpha\alpha'}\, \rho_{\alpha'\beta} =
\int \dd\alpha' \, \Psi^*_\mathrm{qqq}(\alpha')\, \Psi_\mathrm{qqq}(\alpha)
\, \Psi^*_\mathrm{qqq}(\beta)\, \Psi_\mathrm{qqq}(\alpha') =
\Psi^*_\mathrm{qqq}(\beta)\, \Psi_\mathrm{qqq}(\alpha) = \rho_{\alpha\beta}~.
\ee
\\

Reduced density matrices can be constructed by tracing over a subset of
the degrees of freedom, for example
\bea
\rho_{\xi \xi'} = \tr_{\eta,\vec q, \vec Q} \,\, \rho_{\alpha \alpha'}
=
\int
\frac{\dd\eta}{2\eta(1-\eta)}\, \frac{\dd^2 q}{16\pi^3}\, \frac{\dd^2
  Q}{16\pi^3} ~\Psi^*_\mathrm{qqq}(\xi',\eta,\vec
q, \vec Q) \, \Psi_\mathrm{qqq}(\xi,\eta,\vec q, \vec Q)~,
\label{eq:rho_red-xi}
\eea
where $\alpha=\{\xi,\eta,\vec q, \vec Q \}$ and $\alpha' =\{\xi',
\eta,\vec q, \vec Q \}$.  This density matrix describes a mixed state
(except if the ``system'' degree of freedom $\xi$ factorizes from the
$\eta,\vec q, \vec Q$ degrees of freedom of the ``environment'') since
some of the degrees of freedom have been ``traced out''.  Strong
entanglement of $\xi$ with the $\eta,\vec q, \vec Q$ degrees of
freedom will suppress off-diagonal elements of the reduced density
matrix $\rho_{\xi \xi'}$ because the
integral~(\ref{eq:rho_red-xi}) will be small when either
$\xi$ or $\xi'$ is different from its value in the entangled state.  \\

We may also normalize the reduced density matrix as
follows\footnote{We employ a grid of pivot points in the $\xi - \xi'$
  plane such that $\dd\xi = \dd \xi'$ and where the value of $\xi$ at
  the $i^\mathrm{th}$ pivot point is $i/(N_\mathrm{piv}+1)$,
  $i=1,2,\dots,N_\mathrm{piv}$. Hence, $\dd\xi =
  1/(N_\mathrm{piv}+1)$. Also, the Jacobian factor must reduce to
  $1/(2\xi(1-\xi))$ on the $\xi'=\xi$ diagonal, and it must factorize into
  a function of $\xi$ times that same function of $\xi'$, which fixes it to be
$1/\sqrt{2\xi(1-\xi)\,\, 2\xi'(1-\xi')}$.}:
\be \label{eq:rescaled-rho_qqq}
\tilde\rho_{\xi \xi'} = \frac{\dd\xi}
          {\sqrt{2\xi(1-\xi)\,\, 2\xi'(1-\xi')}}~ \rho_{\xi \xi'}~.
\ee
With this normalization the sum of eigenvalues $\lambda_i$ is equal to
1 and we may compute the von~Neumann entanglement entropy
\be
S_\mathrm{vN} = - \sum_i \lambda_i \log \lambda_i ~.
\ee
We use the natural logarithm, so the entropy is measured 
in nats rather than in bits.
\\

The purity $0\le p_\xi \le 1$ of the reduced density matrix is given by
the trace of its square (or by $\sum_i \lambda_i^2$):
\bea
p_\xi &=& \int \frac{\dd\xi}{2\xi(1-\xi)} \int \frac{\dd\xi'}{2\xi'(1-\xi')}
\,\, \rho_{\xi \xi'}\,\, \rho_{\xi' \xi} \\
&= & \int \frac{\dd\xi}{2\xi(1-\xi)} \int \frac{\dd\xi'}{2\xi'(1-\xi')}
\int
\frac{\dd\eta}{2\eta(1-\eta)}\, \frac{\dd^2 q}{16\pi^3}\, \frac{\dd^2
  Q}{16\pi^3}
\int
\frac{\dd\eta'}{2\eta'(1-\eta')}\, \frac{\dd^2 q'}{16\pi^3}\, \frac{\dd^2
  Q'}{16\pi^3} \nn\\
& & ~~~~~~~~~~~~~~~~~~~~
\Psi^*_\mathrm{qqq}(\xi',\eta',\vec q', \vec Q')~
\Psi_\mathrm{qqq}(\xi,\eta',\vec q', \vec Q') ~
\Psi^*_\mathrm{qqq}(\xi,\eta,\vec q, \vec Q)~
\Psi_\mathrm{qqq}(\xi',\eta,\vec q, \vec Q)
~.
\eea
Note that if the $\xi$ degree of freedom factorizes,
$\Psi_\mathrm{qqq}(\xi,\eta,\vec q, \vec Q) = \phi(\xi)\,
\varrho(\eta,\vec q, \vec Q)$, then the purity is $p_\xi=1$. However,
in general such factorization does not occur, e.g.\ due to the ``COM
constraint'' $x_1+x_2+x_3=1$, $\vec k_1+\vec k_2+\vec k_3=0$, or due
to correlations of longitudinal and transverse quark momenta, so
that $p_\xi<1$. In the absence of many-body correlations encoded in the
Hamiltonian we expect that due to the COM constraint alone the impurity
of the reduced density matrix is of order
\be
1- \tr\, \rho^2 = {\cal O}(N_c^{-1})~,
\ee
in the limit of many colors, $N_c\to\infty$.\\~~\\

We can also integrate out the longitudinal degrees of freedom to
construct reduced density matrices over either $\vec q$ or $\vec Q$.
For example,
\be
\rho_{q_x q^{\, \prime}_x} = \tr_{q_y,\xi,\eta, \vec Q} \,\, \rho_{\alpha \alpha'} =
\int \frac{\dd q_y}{16\pi^3}
\frac{\dd\xi}{2\xi(1-\xi)}\,
\frac{\dd\eta}{2\eta(1-\eta)}\,
\frac{\dd^2 Q}{16\pi^3} ~
\Psi^*_\mathrm{qqq}(\xi,\eta,\vec q, \vec Q) \,
\Psi_\mathrm{qqq}(\xi,\eta,\vec q^{\,\prime}, \vec Q)~.
\label{eq:rho_red-q}
\ee
Here, too, in order to obtain a dimensionless density matrix with
properly normalized eigenvalues we should rescale as follows:
\be
\tilde\rho_{q_x q^{\, \prime}_x} = \dd q_x~
\rho_{q_x q^{\, \prime}_x} ~.
\ee
%

\subsection{Numerical estimates for the three quark density matrix}
\label{sec:rho_numerical}

For numerical estimates we employ a simple model for the three quark
wave function $\Psi_\mathrm{qqq}$ due to Schlumpf and
Brodsky~\cite{Schlumpf:1992vq,Brodsky:1994fz}
\begin{equation}
  \Psi_\mathrm{qqq}\left(\left\{x_i,\vec k_i\right\}\right)
  = N_{\mathrm{HO}}
  \, \sqrt{x_1 x_2 x_3}\,\,
  e^{-{\cal M}^2/2\beta^2}
  \label{eq:Psiqqq_HO}
\end{equation}
where ${\cal M}^2 = \sum \frac{\vec k_i^2+m_q^2}{x_i}$ is the
invariant mass squared of the non-interacting three-quark
system~\cite{Bakker:1979eg}.  The normalization of this ``harmonic
oscillator'' wave function follows from eq.~(\ref{eq:Norm_psi3}). The
non-perturbative parameters $m_q=0.26$~GeV and $\beta=0.55$~GeV
have been tuned in Ref.~\cite{Brodsky:1994fz} to the electromagnetic
radius, $R_p=0.76$~fm, the magnetic moments of proton and neutron,
$\mu_{p,n}=2.81/-1.66$, and the axial vector coupling $g_A=1.25$.

The quoted references also present a power-law wave function
\begin{equation}
  \Psi_\mathrm{qqq}\left(\left\{x_i,\vec k_i\right\}\right) = N_{\mathrm{PWR}}
  \,
  \sqrt{x_1 x_2 x_3} \left(1+\frac{{\cal M}^2}{\beta^2}\right)^{-p}~.
  \label{eq:Psiqqq_P}
\end{equation}
The corresponding parameters for this wave function
\eqref{eq:Psiqqq_P} are $p=3.5$, $m=0.263$~GeV,
$\beta=0.607$~GeV~\cite{Brodsky:1994fz}.  Note that this wave function
does not factorize into a product of one quark wave functions, not
even in the absence of the COM constraint.

The only dimensional parameters in the above wave functions are
$\beta^2$ and $m_q^2$. The elements of the normalized density matrix
are dimensionless, so they will only involve the ratio
$m_q^2/\beta^2$. This quantity could also be expressed in terms of the
square of the product of quark mass and proton radius, or as
proton mass times radius~\cite{Schlumpf:1992vq,Brodsky:1994fz}, squared.
\\

The Brodsky-Schlumpf light-front model wave function exhibits
reasonable behavior which is consistent with the empirical knowledge
of the structure of the proton at light-cone momentum fractions $x
\sim 0.1 - 0.5$.  Nevertheless, it would clearly be interesting in the
future to compare to density matrices obtained from three-quark wave
functions $\Psi_\mathrm{qqq}$ which represent solutions of a LF
Hamiltonian with interactions~\cite{Xu:2021wwj,Shuryak:2022thi}.
Also, light-front wave functions at moderate $x$ may become available
from lattice QCD via a large momentum expansion of equal-time
Euclidean correlation functions in instant
quantization~\cite{Ji:2020ect,Ji:2021znw,Liu-Zhao-Schafer-SNOWMASS21}.
Last but not least, the future electron-ion collider EIC will provide
valuable observational constraints on the light-front wave
functions~\cite{Proceedings:2020eah,AbdulKhalek:2021gbh}.  \\

Transforming to unconstrained internal degrees of
freedom~\cite{Bakker:1979eg} we have that $x_1 x_2 x_3 = \eta^2\,
(1-\eta)\, \xi\, (1-\xi)$ and
\be
   {\cal M}^2 = \frac{Q^2}{\eta\, (1-\eta)} + \frac{m_q^2}{1-\eta} +
   \frac{q^2+m_q^2}{\eta\, \xi\, (1-\xi)}~.
\ee

\begin{figure}[htb]
  \includegraphics[width=0.6\textwidth]{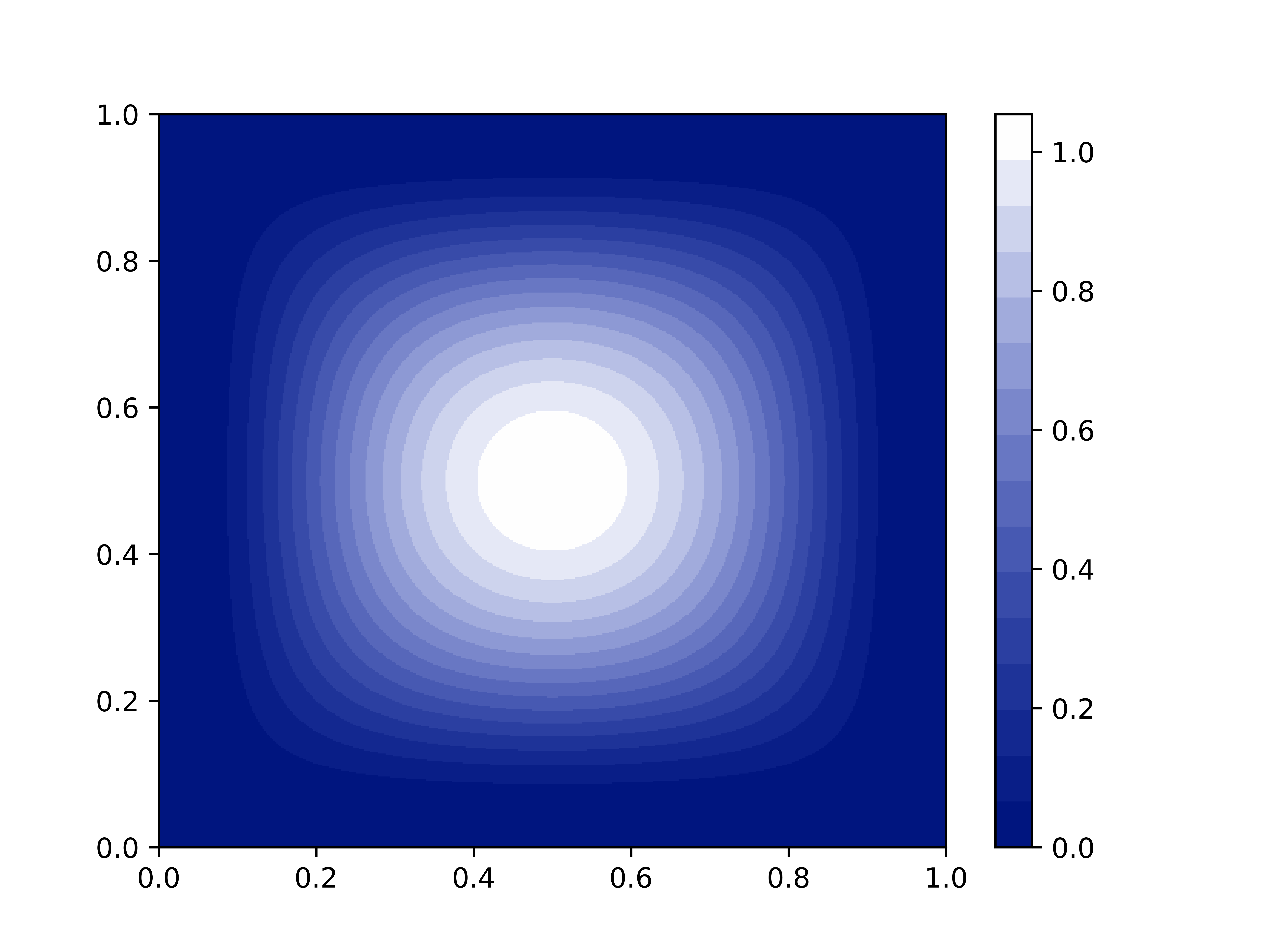}
  \vspace*{-.5cm}
  \caption{\label{fig:qDM-xi} The reduced density matrix
    $\rho_{\xi \xi'}$ for the $|qqq\rangle$
    state ($\Psi_\mathrm{qqq}^\mathrm{HO}$ wave function) plotted over
    $\xi$ vs.\ $\xi'$, both over the interval $(0,1)$. This density
    matrix has a purity of $p_\xi = 0.98$.}
\end{figure}
In fig.~\ref{fig:qDM-xi} we visualize the LO density matrix $\rho_{\xi
  \xi'}$ for the HO wave function. It is clear from the figure that
off-diagonal matrix elements are not strongly suppressed and that this
density matrix represents a nearly separable state. Indeed, we obtain
that the purity is $p_\xi \approx 0.98$, and that the entanglement
entropy is low, $S_\mathrm{vN}\approx 0.06$; analogous values for
other degrees of freedom and wave functions are listed in
table~\ref{tab:purity-wf}.  This indicates that the current model wave
function produces only weak entanglement of spatial degrees of
freedom. In the limit $N_c\to\infty$, with $g^2 N_c$ fixed,
the spatial wave function of the proton factorizes into $N_c$ valence
quark wave functions which are determined by a mean
field~\cite{Witten:1979kh}, where spatial degrees of freedom belonging
to different quarks would not be entangled.
\begin{table}[h!]
  \centering
  \begin{tabular}{ |c|c|c|c|c| }
 \hline
 d.o.f.\ & w.f.\ & purity $\tr\, \rho^2$ & $S_\mathrm{vN}$  \\
 \hline
 $\xi$  & HO  & 0.983 & 0.052 \\ 
 $\xi$  & PWR & 0.992 & 0.029 \\ 
 $\eta$ & HO  & 0.946 & 0.14 \\ 
 $\eta$ & PWR & 0.962 & 0.10 \\
 $Q_x$  & HO  & 0.985 & 0.046 \\
 $Q_x$  & PWR & 0.980 & 0.058 \\
 $q_x$  & HO  & 0.985 & 0.046 \\
 $q_x$  & PWR & 0.980 & 0.058\\
 \hline
\end{tabular}
\caption{Numerical results for the purity and von-Neumann entropy (in
nats) of
  the reduced density matrices over the degree of freedom specified in
  the first column. The three-quark model wave functions are given in
  eqs.~(\ref{eq:Psiqqq_HO}, \ref{eq:Psiqqq_P}),
  respectively~\cite{Schlumpf:1992vq,Brodsky:1994fz}. The numerical
  uncertainty is estimated at a few units on the last quoted digit.}
\label{tab:purity-wf}
\end{table}

\subsection{Color and spin wave function}
\label{sec:color-flavor-spin}

We wrote the symmetric spatial wave function of the proton in
eq.~(\ref{Pstate_short}). That should be multiplied by the color, and
(flavor-)spin wave functions. Let us first restore the anti-symmetric
color space wave function, so now the proton state $|P\rangle =
|P^+,\vec P=0\rangle$ is written as
\be \label{Pstate_color}
|P\rangle = \int [\dd x_i] \int [\dd^2 k_i]\,
\Psi_\mathrm{qqq}\left(\left\{x_i,\vec k_i\right\}\right)\,\,
\frac{1}{\sqrt 6}\sum_{j_1 j_2 j_3} \epsilon_{j_1 j_2 j_3}
\left|\left\{x_i,\vec k_i,j_i\right\}\right>~,
\ee
where $j_1, j_2, j_3 = 1\dots 3$ denote the colors of the quarks.

The three-quark state vectors now include labels for the colors of the
quarks, $\alpha\equiv \left\{x_i,\vec k_i, n_i\right\}$, and
eq.~(\ref{eq:<alpha|P>}) becomes
\bea
\langle\alpha|P\rangle =
\frac{1}{\sqrt 6} \, \epsilon_{n_1 n_2 n_3}
\,\,
\Psi_\mathrm{qqq}\left(\left\{x_i,\vec k_i\right\}\right) \,\,
(2\pi)^3 \,\delta\left(1-\sum_i x_i\right)\,\,
\delta\left(\sum_i \vec k_i\right)~,
\label{eq:<alpha|P>-color}
\eea
Note that $\Psi_\mathrm{qqq}\left(\left\{x_i,\vec k_i\right\}\right)$
is invariant under rotations in color space, therefore it does not
depend on the quark colors $n_i$.

The pure state density matrix $\rho_{\alpha \alpha'}$ from
eqs.~(\ref{eq:Def-rho_pure},\ref{eq:rho-pure}) now reads
\bea \label{eq:rho-pure-color}
\rho_{\alpha \alpha'} =
\frac{1}{6}\, \epsilon_{n_1 n_2 n_3}\, \epsilon_{n_1' n_2' n_3'}
\,\,
\Psi^*_\mathrm{qqq}\left(\left\{x_i',\vec k_i'\right\}\right)~
\Psi_\mathrm{qqq}\left(\left\{x_i,\vec k_i\right\}\right) ~.
\eea
Evidently, this factorizes into a density matrix over color space
times one over momentum space because so does
the pure state~(\ref{Pstate_color}) from which it has been constructed.

When tracing over the degrees of freedom of the ``environment'', we
may now also sum over some or all of the colors using
\bea
\tr_{n_3} &\equiv& \frac{1}{6}\,\sum_{n_3} \epsilon_{n_1 n_2 n_3}\,
\epsilon_{n_1' n_2' n_3}
= \frac{1}{6}\left(\delta_{n_1 n_1'}\delta_{n_2 n_2'}-
\delta_{n_1 n_2'}\delta_{n_2 n_1'}\right)~, \nn\\
\tr_{n_2,n_3} &\equiv& \frac{1}{6}\,\sum_{n_2,n_3} \epsilon_{n_1 n_2 n_3}\,
\epsilon_{n_1' n_2 n_3}
= \frac{1}{3}\,\delta_{n_1 n_1'}~,\nn\\
\tr_{n_1,n_2,n_3} &\equiv& \frac{1}{6}\,\sum_{n_1,n_2,n_3}
\epsilon_{n_1 n_2 n_3}\, \epsilon_{n_1 n_2 n_3}
= 1~.
\eea
Hence, the reduced density matrix shown in previous sections should be
understood as the density matrix obtained after tracing out all quark
colors.

On the other hand, if we trace out all {\em but} one color degree of freedom
then
\bea
\rho_{n n'} = \frac{1}{3}\,\delta_{n n'}~
\int \frac{\dd\xi}{2\xi(1-\xi)} \,
\frac{\dd\eta}{2\eta(1-\eta)}\, \frac{\dd^2 q}{16\pi^3}\, \frac{\dd^2
  Q}{16\pi^3} ~\left| \Psi_\mathrm{qqq}(\xi,\eta,\vec q, \vec Q)\right|^2
= \frac{1}{3}\,\delta_{n n'}~.
\eea
Note that all off-diagonal matrix elements are zero and that this
density matrix is clearly not a separable state. Its purity is
$\sum_{n,n'}\rho_{n n'} \rho_{n' n} = \frac{1}{3}$, i.e.\ the inverse
of the dimension of the fundamental representation of SU(3), which
reflects the entanglement of color. In fact, entanglement is maximal
as all eigenvalues of this reduced density matrix are equal (its
spectrum is degenerate).  For general $N_c$ we have $\rho_{n n'} =
\frac{1}{N_c}\, \delta_{nn'}$, $\tr\, \rho^2 = \frac{1}{N_c}$, and
$S_\mathrm{vN}=\log N_c$.  \\

Lastly, we also multiply by the flavor-spin wave function. The quark-gluon
vertex is flavor independent and we will always trace out flavor degrees
of freedom. Hence we write
\be \label{Pstate_color_spin}
|P\rangle = \int [\dd x_i] \int [\dd^2 k_i]\,
\Psi_\mathrm{qqq}\left(\left\{x_i,\vec k_i\right\}\right)\,\,
\frac{1}{\sqrt 6}\sum_{j_1 j_2 j_3} \epsilon_{j_1 j_2 j_3}
\left|\left\{x_i,\vec k_i,j_i\right\}\right>\,\, |S\rangle~,
\ee
with
\bea
|S\rangle =
\frac{1}{\sqrt{12}}\biggl[\bigl(
2\, \left|\uparrow\uparrow\downarrow\right> -
\left|\downarrow\uparrow\uparrow\right> -
\left|\uparrow\downarrow\uparrow\right>
\bigr) + \bigl(\uparrow\, \leftrightarrow \,\downarrow\bigr)
\biggr]~~~~,~~~~ \langle S|S\rangle = 1~.
\eea
Once again, the LO reduced density matrix shown in sec.~\ref{sec:qqqDM-LO}
should be understood as the density matrix obtained after tracing over
all quark helicities. We refer the reader to ref.~\cite{Beane:2019loz} for
an analysis of entanglement of valence and sea spin in the proton, and
its relation to chiral symmetry breaking.

\subsection{Violation of Bell-CHSH inequality}
\label{sec:Bell-CHSH-color}

In this section we show how color correlations described by
the density operator
\be
\hat\rho = \frac{1}{6}\, \epsilon_{i_1 i_2 i_3}\, \epsilon_{i_1' i_2' i_3'}
\, | i_1, i_2, i_3\rangle\, \langle i_1', i_2', i_3' |
\ee
violate a Bell-CHSH~\cite{Bell:1964kc, Bell:1964fg, Clauser:1969ny}
inequality, indicating that some color correlations are
``stronger than classically possible''. We consider here the simplest
case, a bipartite subsystem of two quarks and ``measurements'' within
a SU(2) subalgebra of color-SU(3). This maps onto the standard system
of two qubits.\\

Consider the Bell-CHSH operator
\be
C_\mathrm{CHSH} = A_1 (B_1+B_2) + A_2 (B_1-B_2)~,
\ee
where the $A_i$ represent the results of two measurements on one part
of the system and $B_i$ the results of independent measurements on
another part. The expectation value of $C_\mathrm{CHSH}$ describes the
correlation of these measurements:
\be
\langle C_\mathrm{CHSH}\rangle = \tr \, C_\mathrm{CHSH}\, \hat\rho =
\frac{1}{6} \, \sum
\epsilon_{i_1 i_2 i_3}\, \epsilon_{i_1' i_2' i_3'}\,
\langle i_1', i_2', i_3' |\, C_\mathrm{CHSH}\, | i_1, i_2, i_3\rangle~.
\ee
The summation in the previous expression is over the quark colors
$i_1, i_2, i_3, i_1', i_2', i_3'$.\\

The measurement operators $A_i, B_i$ act in color subspace 1 and 2,
respectively, and we construct them from the generators of the first
SU(2) subalgebra of color-SU(3), i.e.\ the first three Gell-Mann
matrices $\lambda_1=\sigma_1\oplus0$, $\lambda_2=\sigma_2\oplus0$,
$\lambda_3=\sigma_3\oplus0$ \footnote{$\sigma_i$ are the Pauli
  matrices and $\oplus$ denotes the matrix direct sum $A\oplus B =
  \mathrm{diag}(A,B)$.}.  Similarly, we introduce the identity
corresponding to that subalgebra, $I=1\!\!1_{2\times2}\oplus 0$, as
well as $I_3=1\oplus1\oplus1$.  Hence, our self adjoint measurement
operators are $A_i\otimes I\otimes I_3$ and $I \otimes B_i \otimes
I_3$, and
\be
\langle C_\mathrm{CHSH}\rangle =
\frac{1}{6} \, \sum
\epsilon_{i_1 i_2 3}\, \epsilon_{i_1' i_2' 3}\left[
\left< i_1' |\, A_1\, | i_1\right>\,
\left< i_2' |\, B_1+B_2\, | i_2\right> +
\left< i_1' |\, A_2\, | i_1\right>\,
\left< i_2' |\, B_1-B_2\, | i_2\right> \right]
~.
\ee
.  \\

To obtain a bound on the classical correlation we replace each of the
$A_i$ and $B_i$ by the identity $I$ times one of the eigenvalues of
$\sigma_1$, $\sigma_2$, $\sigma_3$ which are $+1$ or $-1$. 
Therefore, these classical measurements commute. For any
combination of eigenvalues $C_\mathrm{CHSH}$ either takes the value
$\frac{2}{3}$ or $-\frac{2}{3}$. Hence, for any classical probability
distribution of eigenvalues, i.e.\ measurement outcomes, we have that
\be
-\frac{2}{3} \, \le\,  \langle C_\mathrm{CHSH}\rangle_\mathrm{cl} \, \le\,
\frac{2}{3}~.
\ee
\\

The quantum mechanical expectation value of $C_\mathrm{CHSH}$ violates
this bound for certain operators $A_i$, $B_i$. As an example, for
$A_1=\lambda_1$, $A_2=\lambda_3$, $B_1=-(\lambda_1+\lambda_3)/\surd
2$, $B_2=(\lambda_3-\lambda_1)/\surd 2$ we obtain maximal violation,
$\langle C_\mathrm{CHSH}\rangle_q = \frac{2\sqrt{2}}{3}$:
\be
-\frac{2\sqrt{2}}{3} \, \le\,  \langle C_\mathrm{CHSH}\rangle_\mathrm{q}
\, \le\, \frac{2\sqrt{2}}{3}~.
\ee
\\

The violation of Bell-CHSH inequalities indicates that beyond a
classical approximation, an accurate
description of color charge
correlations in the proton at large and moderate $x$ requires
accounting for entanglement and quantum correlations,  (see, also, refs.~\cite{Dumitru:2021tvw,Dumitru:2020fdh}).
In the future, we hope to apply our approach to improve on
classical models of color charge fluctuations in the 
proton at moderately small $x$~\cite{Schlichting:2014ipa,Mantysaari:2016ykx,Mantysaari:2016jaz,Mantysaari:2017dwh,Mantysaari:2018zdd,Mantysaari:2019csc,Mantysaari:2020lhf,Demirci:2021kya},
and to study the sensitivity of specific
observables to quantum color correlations.

\section{Density Matrix for the three quark and one gluon state
  at ${\cal O}(g^2)$}
\label{sec:DM_O-g2}

In this section we consider the emission of a gluon from one of the
quarks.  These corrections give density matrices over the Hilbert
space of three quarks and a gluon. They generate a $|qqqg\rangle$ Fock
state component in the proton, and the corresponding density matrix
$\rho_{qqqg}$.

We need to also consider virtual corrections (see
fig.~\ref{fig:qDM-virt-corr}) due to the exchange of a gluon within
$|P\rangle$ or $\langle P|$.  The next-to-leading order corrections do
not affect $\langle P|P\rangle$ (sec.~III.D in
ref.~\cite{Dumitru:2020gla}). Hence, when we trace $\rho$ over all
degrees of freedom, the contributions from real emissions and virtual
corrections must cancel to restore $\tr\, \rho = \tr\,
\rho_{qqq}^\mathrm{LO} = 1$.  We will check this explicitly below.

\subsection{The density matrix for the four-parton (qqqg) state}

To compute the corrections to the density matrix at order $g^2$ we
begin with the emission of a gluon from the first
quark\footnote{Recall that three momenta are given by $k_1=(x_1 P^+,
  \vec k_1), k_g=(x_g P^+, \vec k_g)$ and so on. From here onward we
  consider a proton with vanishing transverse momentum to make the
  following expressions more compact.}, i.e.\ from the quark with
momentum $k_1$, color $i_1$, and helicity $h_1$:
\bea \label{Pstate_O(g)}
|P^+,\vec P=\vec 0\rangle_{{\cal O}(g)} &=& \int [\dd x_i] \int [\dd^2 k_i]\,
\Psi_\mathrm{qqq}(k_i)\,\,
\frac{1}{\sqrt 6}\sum_{j_1 j_2 j_3} \epsilon_{j_1 j_2 j_3} \nn\\
& &
2g\sum_{\sigma m a} (t^a)_{m j_1} \int\limits_x^{x_1} \frac{\dd x_g}{x_g}
\frac{\dd^2 k_g}{16\pi^3}
\frac{1}{2(x_1-x_g)} \, \sum_{h, h_i}\,
\frac{1}{P^+}\hat\psi_{q\to qg}^{(\sigma h; h_1)}(k_1; k_1-k_g, k_g) \nn\\
& &
\left|m,k_1-k_g,h;\, j_2,k_2,h_2;\, j_3,k_3,h_3\rangle\,
\langle h_1,h_2,h_3|S\rangle~
\otimes |a,k_g,\sigma\right>~.
\eea
Here, a mother quark $\{j_1,k_1,h_1\}$ splits into a daughter quark
$\{m,k_1-k_g,h\}$ and a gluon $\{a,k_g, \sigma\}$, producing a
$|qqqg\rangle$ Fock state in the proton; $x$ is a cutoff on the
light-cone momentum fraction of the gluon required by the soft
singularity in QCD.

The light-cone gauge Fock space amplitude
for the $qg$ state of the quark in $D=4$
dimensions is~\cite{Dumitru:2020gla}
\be
\hat\psi_{q\to qg}^{(\sigma h; h_1)}(p; k_q, k_g) =
\frac{p^+ \sqrt{1- z}}{n^2 + \Delta^2}
\left[(2-z)\, \vec n \cdot \vec \epsilon^*_\sigma
  +i z h \, \vec n \times \vec \epsilon^*_\sigma\right]\,\,
\delta_{h h_1}~,
\ee
where $z=k^+_g/p^+$, $k^+_q/p^+=1-z$, $\vec n = \vec k_g-z \vec p$,
and $\Delta^2=z^2 m^2_\mathrm{col}$ is a ``quark mass'' regulator (in
the light-cone energy denominator) for the collinear DGLAP
singularity; we will sometimes take $m^2_\mathrm{col}\to 0$ when
possible. The cross product in the second term is taken in two
transverse dimensions, $\vec a \times \vec b = \epsilon^{ij}\, a^i\,
b^j$. Also, since we take the mass of the quarks to zero we assume
that their helicity is conserved; therefore, at times we will drop the
superscript $h$ (for the helicity of the daughter quark) on
$\hat\psi$.  There are two more analogous contributions on the
r.h.s.\ of eq.~(\ref{Pstate_O(g)}) corresponding to gluon emission
from quark 2 or quark 3, respectively.\\

Proceeding, we compute the overlap with a prescribed 3-quark, 1-gluon
state. In order to completely characterize such a state we need to also
keep track of which quark $j$ the gluon was emitted from, we do this
explicitly by a superscript:
\be
\alpha^{(j)}_{qqqg} =
\left\{\left\{ n_i, k_i, h_i\right\}, a, k_g, \sigma\right\} ~.
\ee
By analogy to eq.~(\ref{Pstate_O(g)}) we write the corresponding
state vector as
\be
|\alpha^{(j)}\rangle = \left|n_1,k_1-\delta_{j1}k_g,h_1;\,
n_2,k_2-\delta_{j2}k_g,h_2;\, n_3,k_3-\delta_{j3}k_g,h_3\rangle\,
\otimes |a,k_g,\sigma\right>
\ee

We then obtain:
\bea
\left< \alpha^{(j)}_{qqqg}\Big | P\right> &=&
\frac{g}{\sqrt 6}\,\, (2\pi)^3\,\,
\delta\left(1-\sum_i x_i\right)\,\,
\delta\left(\sum_i \vec k_i\right) ~
\langle h_1, h_2, h_3|S\rangle\nn\\
& \sum\limits_m & \left[ \delta_{j1}\, \epsilon_{m n_2 n_3} (t^a)_{n_1 m}
  \,\frac{1}{k_1^+}
  {\hat\psi}_{q\to qg}^{(\sigma h_1)}(k_1; k_1-k_g, k_g) \right.\nn\\
  & & +  \delta_{j2}\, \epsilon_{n_1 m n_3} (t^a)_{n_2 m}
  \,\frac{1}{k_2^+}
  {\hat\psi}_{q\to qg}^{(\sigma h_2)}(k_2; k_2-k_g, k_g) \nn\\
& &  \left.
+  \delta_{j3}\, \epsilon_{n_1 n_2 m} (t^a)_{n_3 m}
  \,\,\frac{1}{k_3^+}
           {\hat\psi}_{q\to qg}^{(\sigma h_3)}(k_3; k_3-k_g, k_g)\right]\,\,
\Psi_\mathrm{qqq}(k_1; k_2; k_3)~.
\label{eq:<qqqg|P>}
\eea
It is clear from this expression that $k_1, k_2, k_3$ denote the
momenta of the {\em parent} quarks so that their longitudinal
(transverse) momenta add to $P^+$ (zero). Also, in each term there is
a $\Theta$-function which ensures that the LC momentum $(x_j-x_g)P^+$
of the daughter quark is positive; we do not write it explicitly.

\begin{figure}[htb]
  \includegraphics[width=0.44\textwidth]{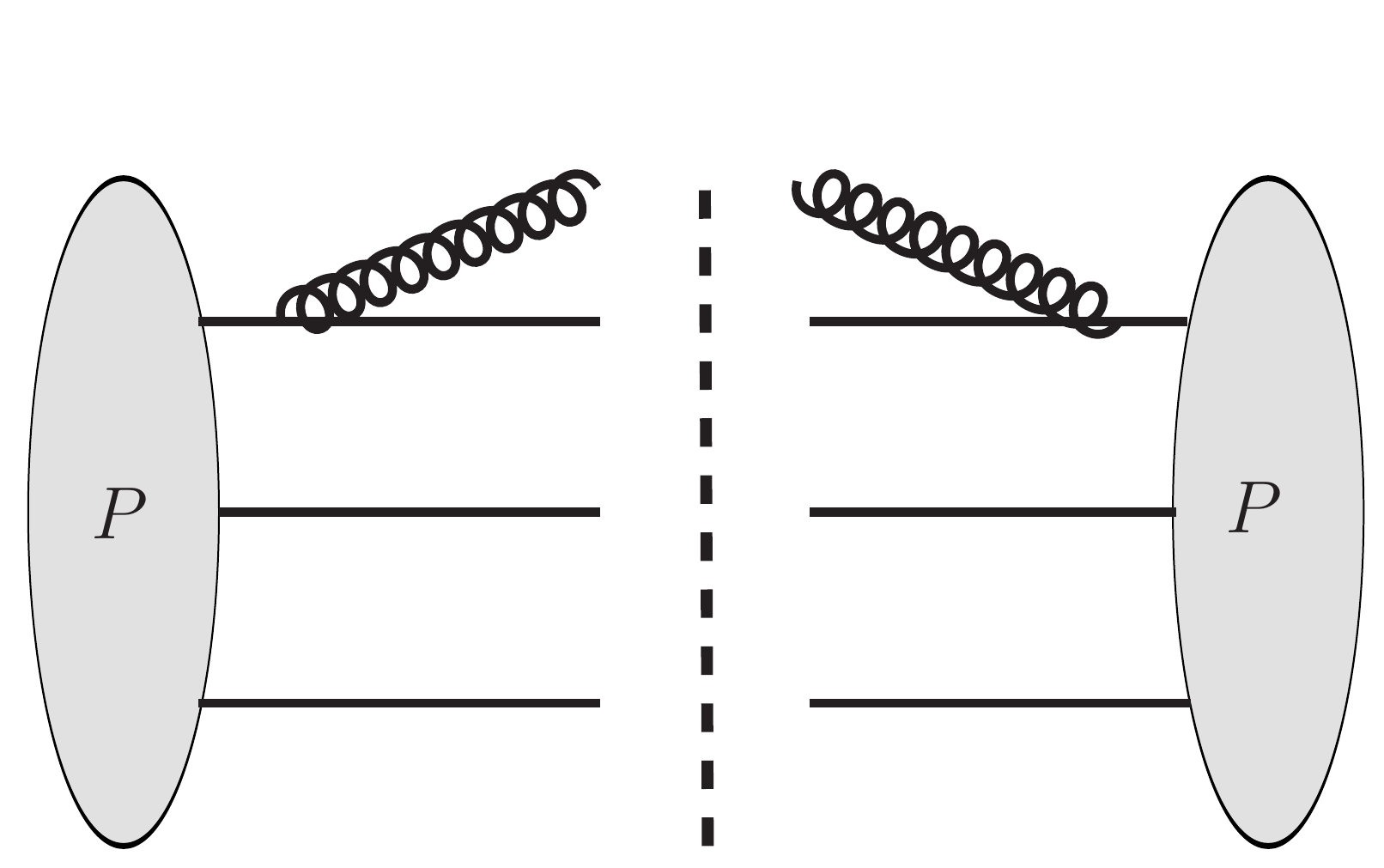}
  \hspace{0.1\textwidth}
  \includegraphics[width=0.44\textwidth]{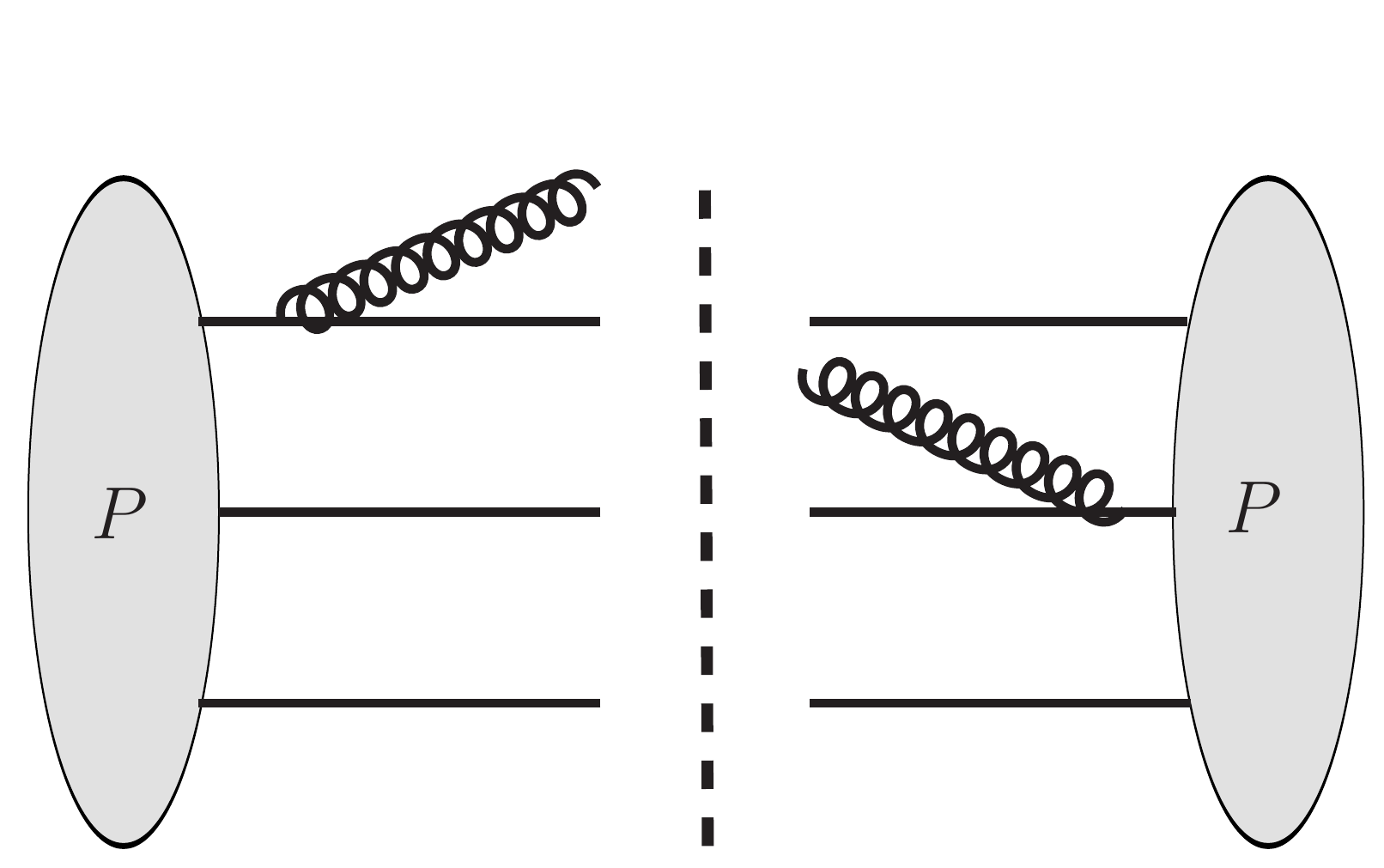}
  \caption{\label{fig:qqqgDM} Density matrices $\rho^{(11')}$ (left) and
    $\rho^{(12')}$ (right) for the three quark plus
    one gluon state $|qqqg\rangle$.}
\end{figure}
Like in eq.~(\ref{eq:Def-rho_pure}), the density matrix describing
3q+1g states is given by the direct product of the previous
expression, as represented in diagram~\ref{fig:qqqgDM}; modulo the
respective COM constraints:
\bea
\rho^{(jj')}_{\alpha \alpha'} ~ (2\pi^3) \,\delta\left(1-\sum_i x_i\right)\,\,
\delta\left(\sum_i \vec k_i\right)~
(2\pi^3) \, \delta\left(1-\sum_i x_i'\right)\,\,
\delta\left(\sum_i \vec k_i'\right)
= \langle\alpha^{(j)}|P\rangle~ \langle P|\alpha^{\prime(j')}\rangle~.
\label{eq:Def-rho_pure-g2}
\eea
Hence, the density matrix for the $|qqqg\rangle$ state is
\bea
\rho^{(jj')}_{\alpha \alpha'} &=&
\frac{g^2}{6}\,\,\langle h_1, h_2, h_3|S\rangle\,
\langle S|h_1', h_2', h_3'\rangle
\sum_{mm'}\nn\\
& &
\left[ \delta_{j'1}\, \epsilon_{m' n_2' n_3'} (t^{a^\prime})_{m' n_1'}
  \,\frac{1}{k_1^{\prime +}}
  {\hat\psi}_{q\to qg}^{*(\sigma^\prime h_1^\prime)}(k_1'; k_1'-k_g', k_g') \right.\nn\\
& & +  \delta_{j'2}\, \epsilon_{n_1' m' n_3'} (t^{a^\prime})_{m' n_2'}
  \,\frac{1}{k_2^{\prime +}}
  {\hat\psi}_{q\to qg}^{*(\sigma' h_2')}(k_2'; k_2'-k_g', k_g') \nn\\
& &  \left.
+  \delta_{j'3}\, \epsilon_{n_1' n_2' m'} (t^{a^\prime})_{m' n_3'}
  \,\,\frac{1}{k_3^{\prime +}}
  {\hat\psi}_{q\to qg}^{*(\sigma' h_3')}(k_3'; k_3'-k_g', k_g')\right] \nn\\
&\times&
\left[ \delta_{j1}\, \epsilon_{m n_2 n_3} (t^a)_{n_1 m}
  \,\frac{1}{k_1^+}
  {\hat\psi}_{q\to qg}^{(\sigma h_1)}(k_1; k_1-k_g, k_g) \right.\nn\\
  & & +  \delta_{j2}\, \epsilon_{n_1 m n_3} (t^a)_{n_2 m}
  \,\frac{1}{k_2^+}
  {\hat\psi}_{q\to qg}^{(\sigma h_2)}(k_2; k_2-k_g, k_g) \nn\\
& &  \left.
+  \delta_{j3}\, \epsilon_{n_1 n_2 m} (t^a)_{n_3 m}
  \,\,\frac{1}{k_3^+}
           {\hat\psi}_{q\to qg}^{(\sigma h_3)}(k_3; k_3-k_g, k_g)\right] \nn\\
& &
\Psi^*_\mathrm{qqq}(k_1'; k_2'; k_3')\,
  \Psi_\mathrm{qqq}(k_1; k_2; k_3)
~.
\eea
Let us consider first the term where the gluon emission occurs from quarks
1 and 1', respectively, i.e.\ $j=j'=1$. 
To make the following expressions more compact we will trace right away
over quark helicities and gluon polarization, so we first compute
\bea
& & \sum_{h_1,h_2,h_3,\sigma}~|\langle h_1, h_2, h_3|S\rangle|^2~
 \frac{1}{k_1^+}\, \frac{1}{k_1^{\prime +}}\,\,
      {\hat\psi}_{q\to qg}^{*(\sigma h_1)}(k_1'; k_1'-k_g', k_g')\,\,
      {\hat\psi}_{q\to qg}^{(\sigma h_1)}(k_1; k_1-k_g, k_g)\nn\\
&=& 2\vec n \cdot \vec n' \, \frac{\sqrt{(1- z)\, (1- z')}}
    {(n^2+\Delta^2)\, (n'^2 + \Delta^{\prime 2})}
\, (2-z-z'+zz')~,
\eea
where $z=x_g/x_1$, $z'=x_g'/x_1'$, $\vec n = \vec k_g - z\vec k_1$,
$\vec n' = \vec k_g' - z'\vec k_1'$, $\Delta^2=z^2 m^2_\mathrm{col}$,
$\Delta^{\prime 2}=z^{\prime 2} m^2_\mathrm{col}$.  With this, the
first contribution to the density matrix for the $|qqqg\rangle$ state
becomes
\bea
\rho^{(11^\prime)}_{\alpha \alpha'} &=& \frac{g^2}{3}
\sum_{mm'} \epsilon_{m' n_2' n_3'} (t^{a^\prime})_{m' n_1'}\,
\epsilon_{m n_2 n_3} (t^a)_{n_1 m} \, \,
\Psi^*_\mathrm{qqq}(k_1'; k_2'; k_3')
 \,
 \Psi_\mathrm{qqq}(k_1; k_2; k_3)
 \nn\\
 & & \Theta(x_1-x_g)\, \Theta(x_1'-x_g')\,\,
 \vec n \cdot \vec n' \, \frac{\sqrt{(1- z)\, (1- z')}}
     {(n^2+\Delta^2)\, (n'^2 + \Delta^{\prime 2})}
 \, (2-z-z'+zz') ~.
\label{eq:rho_qqqg_11'}
\eea
(Here, the matrix indices $\alpha, \alpha'$ exclude quark helicities
and gluon polarization.)
There are two more analogous contributions corresponding to gluon emission
from quarks 2,2' or from quarks 3,3', respectively.
\\~~\\~~\\

Now we derive the density matrix for the case where the gluon in
$|P\rangle$ is emitted by quark 1 while that in $\langle P|$ is
emitted by quark 2. Using
\bea
\sum_{h_1,h_2,h_3}~|\langle h_1, h_2, h_3|S\rangle|^2~h_1h_2 &=&
  \frac{1}{3}~,\\
\sum_{h_1,h_2,h_3}~|\langle h_1, h_2, h_3|S\rangle|^2~h_1h_3 &=&
  -\frac{2}{3}~,\\
\sum_{h_1,h_2,h_3}~|\langle h_1, h_2, h_3|S\rangle|^2~h_2h_3 &=&
  -\frac{2}{3}~,
\eea
we can use the following expression to trace over quark helicities and
gluon polarization:
\bea
& & \sum_{h_1,h_2,h_3,\sigma}~|\langle h_1, h_2, h_3|S\rangle|^2~
\frac{1}{k_1^+\, k_2^{\prime +}}\,
{\hat\psi}_{q\to qg}^{(\sigma h_2)\, *}(k_2'; k_2'-k_g', k_g')\,\,
{\hat\psi}_{q\to qg}^{(\sigma h_1)}(k_1; k_1-k_g, k_g) \nn\\
&=& 2\vec n \cdot \vec n' \,
\frac{\sqrt{(1- z)\, (1- z')}}
     {(n^2+\Delta^2)\, (n'^2 + \Delta^{\prime 2})}
\, (2-z-z'+\frac{1}{2}zz'(1+\langle h_1 h_2\rangle))~,
\eea
where $z=x_g/x_1$, $z'=x_g'/x_2'$, $\vec n = \vec k_g - z\vec k_1$,
$\vec n' = \vec k_g' - z'\vec k_2'$, $\Delta^2 =z^2m^2_\mathrm{col}$,
$\Delta^{\prime 2}=z^{\prime 2} m^2_\mathrm{col}$.  Then,
\bea
\rho^{(12^\prime)}_{\alpha \alpha'} &=& \frac{g^2}{3}\,
\sum_{mm'} \epsilon_{n_1' m' n_3'} (t^{a^\prime})_{m' n_2'}\,
\epsilon_{m n_2 n_3} (t^a)_{n_1 m} \, \,
\Psi^*_\mathrm{qqq}(k_1'; k_2'; k_3')\,\,
\Psi_\mathrm{qqq}(k_1; k_2; k_3)\nn\\
& & \Theta(x_1-x_g)\, \Theta(x_2'-x_g')\,\,
\vec n \cdot \vec n' \,
\frac{\sqrt{(1- z)\, (1- z')}}
     {(n^2+\Delta^2)\, (n'^2 + \Delta^{\prime 2})}
\, (2-z-z'+\frac{1}{2}zz'(1+\langle h_1 h_2\rangle)) 
~.
\label{eq:rho_qqqg_12'}
\eea
There are five more analogous contributions corresponding to
emission from quark pairs (13'), (21'), (23'), (31'), (32')respectively.

\subsection{Trace of $\rho_{qqqg}$}

As a first step we need to determine the integration measure over the
gluon (spatial) degrees of freedom. With
\bea
\langle\alpha^{\prime (1')} | \alpha^{(1)}\rangle &=& \langle
k_1'-k_g'; k_2'; k_3' | k_1-k_g; k_2; k_3\rangle\,\, \langle k_g' | k_g\rangle
\nn\\
&=& (16\pi^3)^4\, k_g^{\prime +}\, \delta(k_g'-k_g)
\, (k_1^{\prime +}-k_g^{\prime +})\, \delta(k_1'-k_g'-k_1+k_g)
\, k_2^{\prime +}\, \delta(k_2'-k_2)
\, k_3^{\prime +}\, \delta(k_3'-k_3) \\
&=& (16\pi^3)^4\, k_g^+\, \delta(k_g'-k_g)
\, (k_1^+-k_g^+)\, \delta(k_1'-k_g'-k_1+k_g)
\, k_2^+\, \delta(k_2'-k_2)
\, k_3^+\, \delta(k_3'-k_3)
\label{eq:<alpha^{'(1')}|alpha^{(1)}>}
\eea
we define
\be  \label{eq:dalpha1'}
\dd\alpha^{\prime (1')} =
\frac{1}{2}\frac{\dd x_1'\dd x_2'\dd x_3'}{8\, x_1'\, x_2'\, x_3'}
\,\delta(1-\sum x_i')\, \frac{\dd^2 k_1'\,\dd^2 k_2'\,\dd^2 k_3'}{(2\pi)^6}
\,\delta(\sum \vec k_i')\,
\frac{\dd x_g'}{x_g'}\frac{\dd^2 k_g'}{16\pi^3}\, \frac{x_1'}{x_1'-x_g'}
\ee
and
\be  \label{eq:dalpha1}
\dd\alpha^{(1)} =
\frac{1}{2}\frac{\dd x_1\dd x_2\dd x_3}{8\, x_1\, x_2\, x_3}
\,\delta(1-\sum x_i)\, \frac{\dd^2 k_1\,\dd^2 k_2\,\dd^2 k_3}{(2\pi)^6}
\,\delta(\sum \vec k_i)\,
\frac{\dd x_g}{x_g}\frac{\dd^2 k_g}{16\pi^3}\, \frac{x_1}{x_1-x_g}
\ee
so that
\bea
\int\dd\alpha^{\prime (1')}\, \langle\alpha^{\prime (1')} | \alpha^{(1)}\rangle
&=& \frac{1}{2} \, (2\pi)^3\, \delta(1-\sum x_i)\, \delta(\sum \vec k_i)
~, \\
\int\dd\alpha^{(1)}\, \langle\alpha^{\prime (1')} | \alpha^{(1)}\rangle
&=& \frac{1}{2} \, (2\pi)^3\, \delta(1-\sum x_i')\, \delta(\sum \vec k_i')
\eea
which is analogous to eq.~(\ref{eq:<alpha'|alpha>-qqq}).

Similarly, with
\bea
\langle\alpha^{\prime (2')} | \alpha^{(1)}\rangle &=& \langle
k_1'; k_2'-k_g'; k_3' \,|\, k_1-k_g; k_2; k_3\rangle\,\,
\langle k_g' | k_g\rangle
\nn\\
&=& (16\pi^3)^4\, k_g^+\, \delta(k_g'-k_g)
\, k_1^{\prime +}\, \delta(k_1'-k_1+k_g)
\, k_2^{+}\, \delta(k_2'-k_g-k_2)
\, k_3^+\, \delta(k_3'-k_3)
\label{eq:<alpha^{'(2')}|alpha^{(1)}>}
\eea
and
\be  \label{eq:dalpha2'}
\dd\alpha^{\prime (2')} =
\frac{1}{2}\frac{\dd x_1'\dd x_2'\dd x_3'}{8\, x_1'\, x_2'\, x_3'}
\,\delta(1-\sum x_i')\, \frac{\dd^2 k_1'\,\dd^2 k_2'\,\dd^2 k_3'}{(2\pi)^6}
\,\delta(\sum \vec k_i')\,
\frac{\dd x_g'}{x_g'}\frac{\dd^2 k_g'}{16\pi^3}\, \frac{x_2'}{x_2'-x_g'}
\ee
we obtain
\bea
\int\dd\alpha^{(1)}\, \langle\alpha^{\prime (2')} | \alpha^{(1)}\rangle
&=& \frac{1}{2} \, (2\pi)^3\, \delta(1-\sum x_i')\, \delta(\sum \vec k_i')
~, \\
\int\dd\alpha^{\prime (2')}\, \langle\alpha^{\prime (2')} | \alpha^{(1)}\rangle
&=& \frac{1}{2} \, (2\pi)^3\, \delta(1-\sum x_i)\, \delta(\sum \vec k_i)
~.
\eea
\\

The trace over $\rho^{(jj^\prime)}_{\alpha \alpha'}$ is again given by
the expression in eq.~(\ref{eq:tr_rho^}) on the left,
\bea
\frac{1}{\frac{1}{4}\langle P|P\rangle} \int\dd\alpha^{(j)}\,
\int\dd\alpha^{\prime (j')}\, \langle\alpha^{\prime (j')} | \alpha^{(j)}\rangle
\, \rho^{(jj^\prime)}_{\alpha \alpha'}
= \int\dd\alpha^{(j)}\, \rho^{(jj^\prime)}_{\alpha \alpha'} ~.
\eea
The expression on the left is analogous to the trace of a matrix,
$\sum_{i,j}\, M_{ij} \, \langle i|j\rangle$ where $\langle i|j\rangle
= \delta_{ij}$.  On the r.h.s.\ of the above, in
$\rho^{(jj^\prime)}_{\alpha \alpha'}$ the primed quark and gluon
momenta have to be expressed in terms of the unprimed momenta as
determined by the $\delta$-functions in
eqs.~(\ref{eq:<alpha^{'(1')}|alpha^{(1)}>},
\ref{eq:<alpha^{'(2')}|alpha^{(1)}>}), respectively.  \\

We first trace eq.~(\ref{eq:rho_qqqg_11'}) over the quark
degrees of freedom by summing over their colors, and integrating over
their longitudinal and transverse momenta.  This leads to the reduced
gluon density matrix
\bea
\rho^{(j=j^\prime)}_{\alpha \alpha'} &=& 2g^2\, \tr\, t^{a'} t^a\,
\frac{1}{2}\int[\dd x_i]\int[\dd^2k_i]\, \Theta(x_1-x_g)\,
\Theta(x_1-x_g')\, \Psi^*_\mathrm{qqq}(k_1-k_g+k_g'; k_2; k_3)\,
\Psi_\mathrm{qqq}(k_1; k_2; k_3)
\nn\\
& &~~~
\vec n \cdot \vec n' \, \frac{\sqrt{(1- z)\, (1- z')}}
     {(n^2+\Delta^2)\, (n'^2 + \Delta^{\prime 2})}
 \, (2- z - z'+zz')
     ~,
\label{eq:tr-qqq_rho_qqqg_11'}
\eea
where now $z=x_g/x_1$, $z'=x_g'/(x_1-x_g+x_g')$, $\vec n = \vec k_g -
z\vec k_1$, $\vec n' = \vec k_g' - z'(\vec k_1-\vec k_g + \vec k_g')$,
$\Delta^2=z^2m^2_\mathrm{col}$, $\Delta^{\prime 2} =z^{\prime
  2}m^2_\mathrm{col}$.  The indices $\alpha$ now refer exclusively to
the color and longitudinal and transverse momentum of the gluon. This
expression includes a factor of 3 to account for the diagrams where
the gluon emission occurs from quarks 2,2' or 3,3', respectively (this
uses the symmetry of the three quark wave function under exchange of
any two quarks).  \\

As a final step, we trace out the gluon by summing over $a'=a$ and
integrating over
\be  \label{eq:g-measure_x1}
\int\limits_x^1\frac{\dd x_g}{x_g}\int\frac{\dd^2 k_g}{16\pi^3}\,
\frac{x_1}{x_1-x_g}~.
\ee
Here we encounter a UV divergent integral over the transverse momentum
$\vec k_g$ of the gluon.  To regularize this quantity we subtract a UV
contribution which we shall add back to the ${\cal O}(g^2)$ virtual
correction to $\rho^\mathrm{LO}$ in the following
sec.~\ref{sec:qqqDM-NLO}. It is given by the $\Lambda$-dependent part
of the integrand of the ${\cal O}(g^2)$ contribution to the wave
function renormalization factor
\bea
& &
\sqrt{Z_q(x_1)\,Z_q(x_1')\,Z_q(x_2)\,Z_q(x_2')\,Z_q(x_3)\,Z_q(x_3')\,} - 1 =
\nn\\
& &
-\frac{1}{2}C_q(x_1;\frac{\Lambda}{m_\mathrm{col}})-
\frac{1}{2}C_q(x_1';\frac{\Lambda}{m_\mathrm{col}})
-\frac{1}{2}C_q(x_2;\frac{\Lambda}{m_\mathrm{col}})
-\frac{1}{2}C_q(x_2';\frac{\Lambda}{m_\mathrm{col}})
-\frac{1}{2}C_q(x_3;\frac{\Lambda}{m_\mathrm{col}})
-\frac{1}{2}C_q(x_3';\frac{\Lambda}{m_\mathrm{col}})~,
\eea
which is written explicitly in eqs.~(\ref{eq:Cq_x1_Lambda}, \ref{eq:1-6Cq}).
This gives the regularized trace
\bea
\tr\, \rho^{(j=j^\prime)} &=& 8g^2\,
\frac{1}{2}\int[\dd x_i]\int[\dd^2k_i]
\int\limits_x^{x_1}\frac{\dd x_g}{x_g}
\left|\Psi_\mathrm{qqq}(k_1; k_2; k_3)\right|^2
     \left[1+\left(1-\frac{x_g}{x_1}\right)^2\right]\,
 \int\frac{\dd^2 k_g}{16\pi^3}\,\,
     \left[\frac{1}{k_g^2+\Delta^2}
       - \frac{1}{k_g^2+\Lambda^2}\right]~,
\eea
where $\Lambda=(x_g/x_1) M_\mathrm{UV}$ while $\Delta=(x_g/x_1) m_\mathrm{col}$.
If the integral over $x_g$ is dominated by $x_g$ much less than
typical quark light-cone momentum fractions then we can replace the
upper limit by $\langle x_q\rangle$ and use the normalization
condition~(\ref{eq:Norm_psi3}) for the three-quark wave function to
simplify further:
\bea
\simeq \frac{g^2}{\pi^2}\,\log\frac{\langle x_q\rangle}{x}\,
     \log \frac{M^2_\mathrm{UV}}{m^2_\mathrm{col}}    ~.
\label{eq:tr-gqqq_rho_qqqg_11'_reg}
\eea
The previous expressions exhibit a dependence on the IR cutoffs, $x$
for the soft singularity and $m_\mathrm{col}$ for the collinear
singularity, and on the UV regulator $M_\mathrm{UV}$ in $D=4$
dimensions. A dependence of the entanglement entropy on the logarithm of
the UV cutoff has also been found in ref.~\cite{Kovner:2015hga}.\\~~\\

Similarly, to trace eq.~(\ref{eq:rho_qqqg_12'}) over quark degrees of
freedom we set $k_1-k_g=k_1'$, $k_2=k_2'-k_g'$, $k_3=k_3'$, $n_i'=n_i$:
\bea
\rho^{(12^\prime)}_{\alpha \alpha} &=& -\frac{g^2}{3}\, \tr \, t^{a'} t^a\,
\frac{1}{2}\int[\dd x_i]\int[\dd^2k_i] \,\,
\vec n \cdot \vec n' \, \frac{\sqrt{(1- z)\, (1- z')}}
     {(n^2+\Delta^2)\, (n'^2 + \Delta^{\prime 2})}
\, (2-z-z'+\frac{1}{2}zz'(1+\langle h_1h_2\rangle))\nn\\
& &
\Theta(x_1-x_g)\,\Theta(1-(x_2+x_g'))\, 
\Psi^*_\mathrm{qqq}(k_1-k_g; k_2+k_g'; k_3)\,
\Psi_\mathrm{qqq}(k_1; k_2; k_3)\,
~,
\label{eq:tr-qqq_rho_qqqg_12'}
\eea
with $z=x_g/x_1$, $z'=x_g'/(x_2+x_g')$, $\vec n = \vec k_g - z\vec
k_1$, $\vec n' = \vec k_g' - z'(\vec k_2+\vec k_g')$, $\Delta=z
m_\mathrm{col}$, $\Delta' =z' m_\mathrm{col}$.  The trace over the
remaining gluon degrees of freedom is obtained by summing over $a'=a$,
setting $k_g=k_g'$, and integrating with the measure
\be
\int\frac{\dd x_g}{x_g}\int\frac{\dd^2 k_g}{16\pi^3}\,
\frac{x_1}{x_1-x_g}~.
\ee
Hence,
\bea
\tr\, \rho^{(12^\prime)} &=& -\frac{2g^2}{3}\,
\int\frac{\dd x_g}{x_g}\int\frac{\dd^2 k_g}{16\pi^3}
\int[\dd x_i]\int[\dd^2k_i] \,\, \Theta(x_1-x_g)\, \Theta(1-x_2-x_g)\nn\\
& &
\Psi^*_\mathrm{qqq}(k_1-k_g; k_2+k_g; k_3)\,\,
\Psi_\mathrm{qqq}(k_1; k_2; k_3) \nn\\
& &
\vec n \cdot \vec n' \, \sqrt\frac{x_1\, x_2}{(x_2+x_g)\,(x_1-x_g)} \, \frac{1}
     {(n^2+\Delta^2)\, (n'^2 + \Delta^{\prime 2})}
\, (2-z-z'+\frac{1}{2}zz'(1+\langle h_1h_2\rangle))
~.
\label{eq:tr-gqqq_rho_qqqg_12'_shift}
\eea
Note that the integral over $\vec k_g$ converges in the UV because
some of the transverse momentum arguments of $\Psi^*_\mathrm{qqq}$ are
shifted by $\pm \vec k_g$.  This expression cancels against
eq.~(\ref{eq:tr_rho12_virt}).\\

\subsection{${\cal O}(g^2)$ virtual correction to the three-quark
  density matrix}
\label{sec:qqqDM-NLO}

\begin{figure}[htb]
  \includegraphics[width=0.44\textwidth]{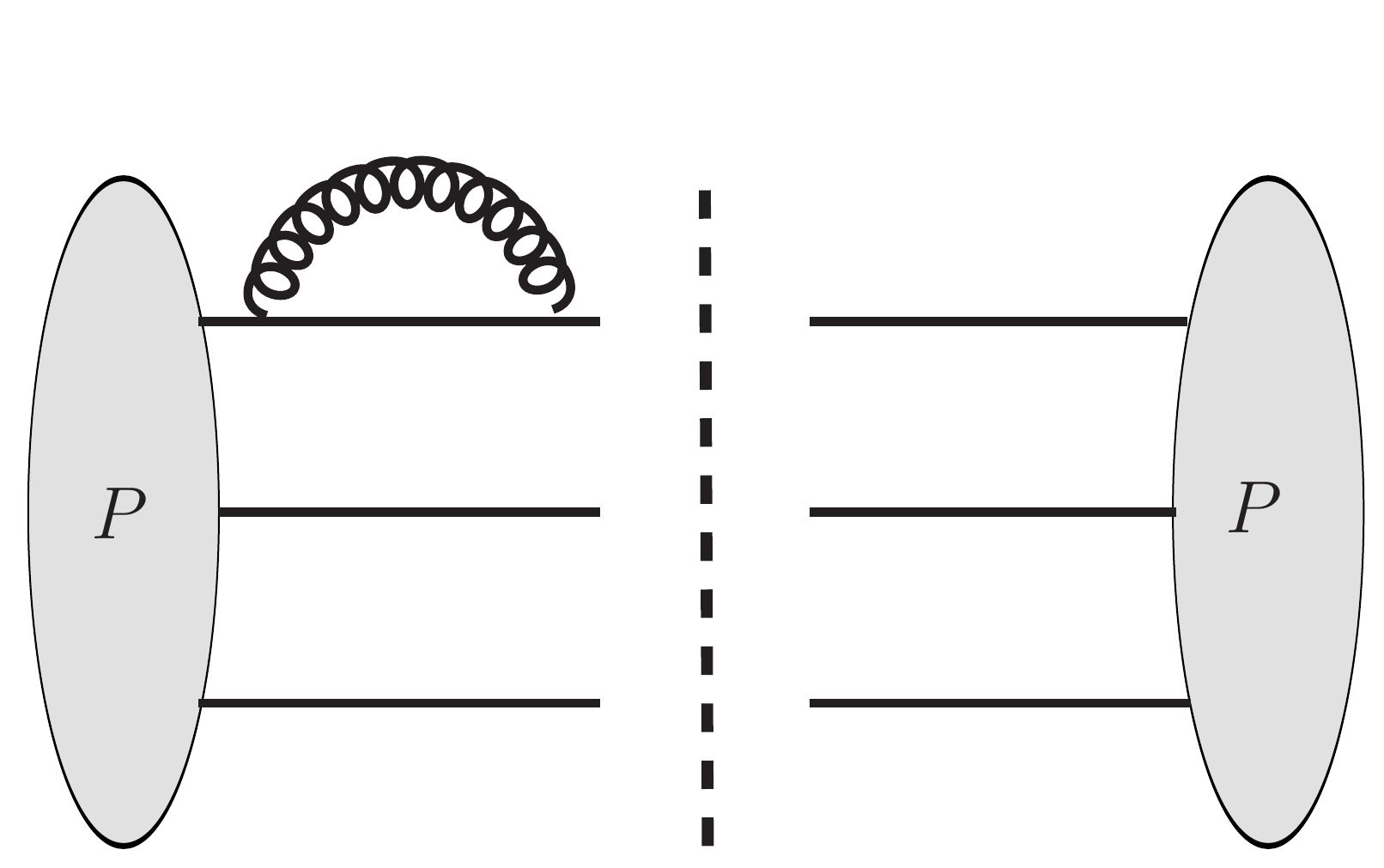}
  \hspace{0.1\textwidth}
  \includegraphics[width=0.44\textwidth]{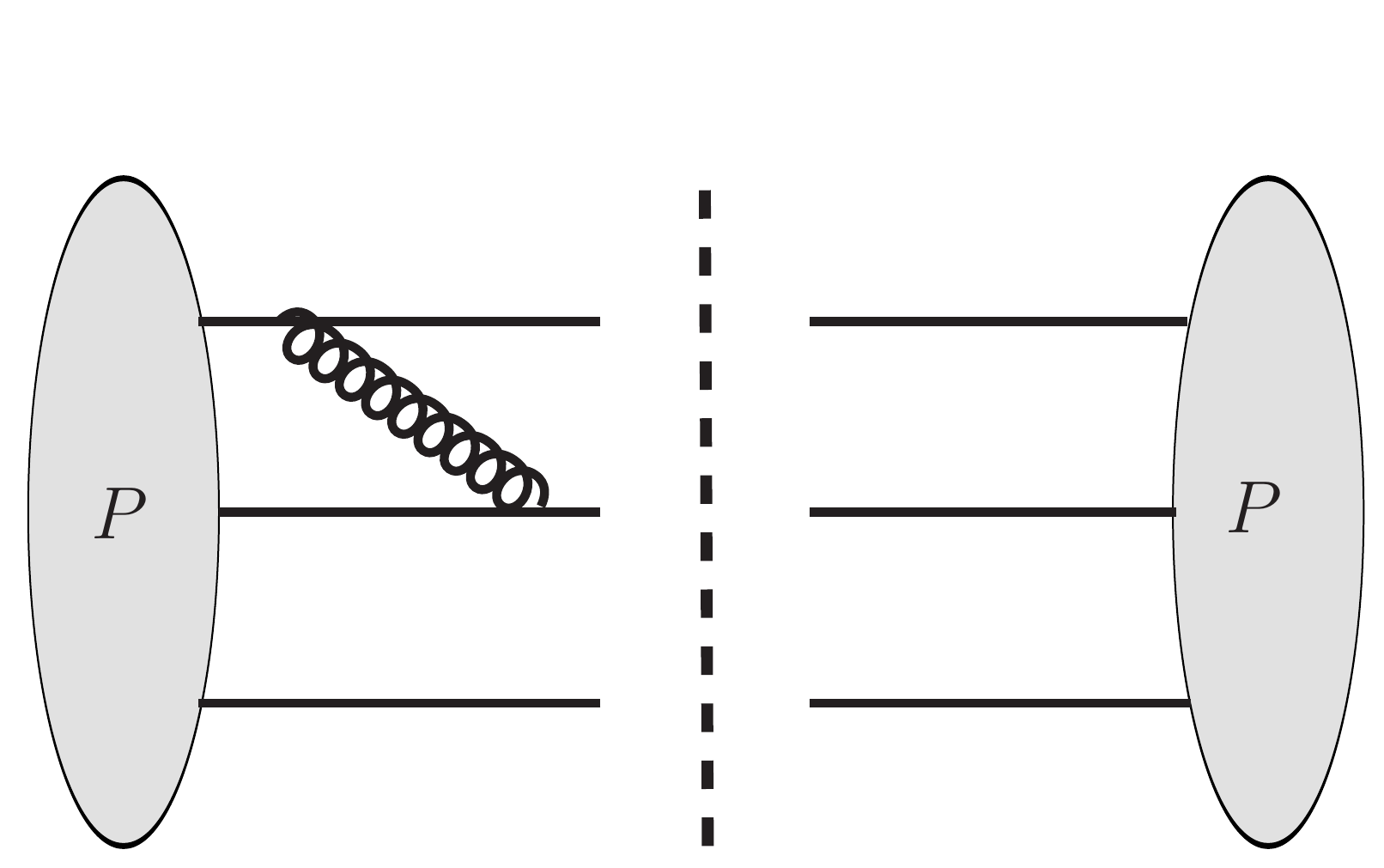}
  \caption{\label{fig:qDM-virt-corr} Virtual corrections
    to the three quark density matrix.}
\end{figure}
In this section we derive the corrections to the LO three-quark
density matrix from sec.~\ref{sec:qqqDM-LO}. These are due to i) the
emission and reabsorption of a gluon by a quark, and ii) the exchange
of a gluon by two distinct quarks, either in $|P\rangle$ or in
$\langle P|$; see fig.~\ref{fig:qDM-virt-corr}.

We begin with the former correction which amounts to multiplying each
quark state vector in eq.~(\ref{Pstate_short}) or (\ref{Pstate_color})
by a wave function renormalization factor $Z_q^{1/2}(x_i)$. The factor
$Z_q(x_i)$ equals 1 minus the ${\cal O}(g^2)$
correction~\cite{Dumitru:2020gla}
\bea
C_q(x_i) = \frac{1}{2x_i}\int \frac{\dd x_g}{2x_g}\frac{\dd^2 k_g}{(2\pi)^3}
\frac{1}{2x_q}\, \sum_{n,a,\sigma}
\left| \psi_{q\to qg}^{(h\sigma)}(p; k_q, k_g)/P^+
\right|^2~.
\eea
The three-momenta satisfy $p=k_q+k_g$; $a, n$ are the colors of the
daughter gluon and quark, respectively; and $\sigma$ denotes the
polarization of the gluon.

The quark wave function renormalization factor is UV divergent:
\bea
C_q(x_1) &=&
\frac{g^2C_F}{8\pi^2} \int_{x/x_1}^{1}
\frac{\dd z}{z}\biggl [1 + (1-z)^2 \biggr ]\, A_0(z m_\mathrm{col})
\label{eq:Cq}\\
A_0(\Delta) &=& 4\pi \int \frac{\dd^2n}{(2\pi)^2}
\frac{1}{\vec{n}^2 + \Delta^2}~.  \label{eq:A0_div}
\eea
$m_\mathrm{col}$ denotes a ``quark mass like'' regulator of the
collinear singularity.  One may use dimensional
regularization~\cite{Beuf:2016wdz,Hanninen:2017ddy,Dumitru:2020gla} to
regularize $C_q(x_1)$. Here, instead we employ a scheme where we
subtract the contribution from a mass scale in the ultraviolet,
$\Lambda^2 = z^2 M^2_\mathrm{UV}$.  We define the regularized function
\bea
A_0^{\mathrm{reg}}(\Lambda/\Delta) &=& A_0(\Delta) - A_0(\Lambda) =
4\pi \int \frac{\dd^2n}{(2\pi)^2}
\left[\frac{1}{\vec{n}^2 + \Delta^2} - \frac{1}{\vec{n}^2 + \Lambda^2} \right]
= \log\frac{\Lambda^2}{\Delta^2}~.
\label{eq:A0_reg}
\eea
This function is now used in eq.~(\ref{eq:Cq}) to obtain
\bea
C_q(x_1;\, x,\frac{M_\mathrm{UV}}{m_\mathrm{col}}) &=&
2g^2C_F \int_{x}^{x_1}
\frac{\dd x_g}{x_g}\int \frac{\dd^2n}{16\pi^3}\, \frac{x_1}{x_1-x_g} \nn\\
& &\frac{x_1-x_g}{x_1}
\biggl [1 + (1-\frac{x_g}{x_1})^2 \biggr ]
\left[\frac{1}{\vec{n}^2 + \Delta^2} - \frac{1}{\vec{n}^2 + \Lambda^2}\right]
~, \label{eq:Cq_x1_Lambda}
\eea
where $\Delta^2 = x_g^2 m^2_\mathrm{col}/x_1^2$, $\Lambda^2 = x_g^2
M^2_\mathrm{UV}/x_1^2$.  In effect we have added to $-C_q(x_1)$ the
infinite term
\be
\frac{g^2 C_F}{2\pi} \int_{x/x_1}^1 \frac{\dd z}{z} \,
\left[1+(1-z)^2\right]\, \int\frac{\dd^2 n}{(2\pi)^2}
\frac{1}{\vec n^2 + \Lambda^2}
~.    \label{eq:Cq-Lambda}
\ee
This is the same UV divergent contribution we previously subtracted from
$\tr\, \rho^{(11')}$.
\\

In all, the first ${\cal O}(g^2)$ correction to the three-quark
density matrix amounts to multiplying
eqs.~(\ref{eq:Def-rho_pure}, \ref{eq:rho-pure}) by
\bea
& &
\sqrt{Z_q(x_1)\, Z_q(x_2)\, Z_q(x_3)\, Z_q(x_1')\, Z_q(x_2')\, Z_q(x_3')\, }
\nn\\
&=&
1 - \frac{C_q(x_1;\, x,\frac{M_\mathrm{UV}}{m_\mathrm{col}})}{2}
- \frac{C_q(x_2;\, x,\frac{M_\mathrm{UV}}{m_\mathrm{col}})}{2}
- \frac{C_q(x_3;\, x,\frac{M_\mathrm{UV}}{m_\mathrm{col}})}{2}
- \frac{C_q(x_1';\, x,\frac{M_\mathrm{UV}}{m_\mathrm{col}})}{2}
- \frac{C_q(x_2';\, x,\frac{M_\mathrm{UV}}{m_\mathrm{col}})}{2}
- \frac{C_q(x_3';\, x,\frac{M_\mathrm{UV}}{m_\mathrm{col}})}{2}~.
\label{eq:1-6Cq}
\eea
Here, we discard contributions beyond ${\cal O}(g^2)$.
This generates a correction factor for the
trace of the three-quark density matrix:
\bea
1 - 3\,\, \frac{1}{2} \int[\dd x_i]\int[\dd^2k_i] \,\,
\left|\Psi_\mathrm{qqq}(k_1; k_2;
k_3)\right|^2\, \, C_q(x_1;\, x,\frac{M_\mathrm{UV}}{m_\mathrm{col}})~.
\label{eq:1-3Int-Cq}
\eea
The ${\cal O}(g^2)$ correction cancels against the contribution from
the trace of the density matrix for the $|qqqg\rangle$ state,
eq.~(\ref{eq:tr-gqqq_rho_qqqg_11'_reg}).  \\~~\\

We now move on to the second kind of ${\cal O}(g^2)$ correction due to
the exchange of a gluon by two quarks. Let quark 1 emit and
quark 2 absorb the gluon in $|P\rangle$:
\bea
|P^+,\vec P=0\rangle_{{\cal O}(g^2)} &=& \int [\dd x_i] \int [\dd^2 k_i]\,
\Psi_\mathrm{qqq}\left(k_1; k_2; k_3)\right)\,\,
\frac{1}{\sqrt 6}\sum_{j_1 j_2 j_3} \epsilon_{j_1 j_2 j_3} \sum_{h_i,h,h'}
\langle h_1, h_2, h_3\, |\, S\rangle  \nn\\
& &
4g^2\sum_{\sigma, a, n, m} (t^a)_{m j_1} (t^a)_{n j_2}
\int \frac{\dd x_g}{x_g}
\frac{\dd^2 k_g}{16\pi^3}
\frac{\Theta(\mathrm{min}(x_1,1-x_2)-x_g)}{2(x_1-x_g)} \,
\frac{1}{P^+}\hat\psi_{q\to qg}^{(\sigma h; h_1)}(k_1; k_1-k_g, k_g)
\nn\\
& &
\frac{1}{2(x_2+x_g)} \,
\frac{1}{P^+}\hat\psi_{qg\to q}^{(\sigma h'; h_2)}(k_2, k_g; k_2+k_g)
\,
\left|m,k_1-k_g,h;\, n,k_2+k_g,h';\, j_3,k_3,h_3\right>
~.      \label{eq:Pstate_O(g2Xchange)}
\eea
Here, the light-cone amplitude in $D=4$ dimensions for the absorption
of a gluon by a quark is~\cite{Dumitru:2020gla}
\be
\hat\psi_{qg\to q}^{(\sigma h; h_1)}(k_q, k_g; p) = -
\frac{p^+ \sqrt{1-z}}{n^2+\Delta}
\left[(2-z)\, \vec n \cdot \vec \epsilon_\sigma
  -i z h \, \vec n \times \vec \epsilon_\sigma\right]\,
\delta_{h h_1}~,
\ee
where again $z=k_g^+/p^+$ and $\vec n=\vec k_g - z\vec p$.

We compute the overlap with a prescribed 3-quark state:
\bea
\left< \left\{ n_i, k_i, h_i\right\}
\Big | P\right> &=&
(2\pi)^3\, \delta(1-x_1-x_2-x_3)\, \delta(\vec k_1 + \vec k_2 + \vec k_3)
\,\, \langle h_1,h_2,h_3|S\rangle\, \sum_{j_1. j_2}
\nn\\
& & g^2\sum_{\sigma, a} (t^a)_{n_1 j_1} (t^a)_{n_2 j_2}
\int \frac{\dd x_g}{x_g}
\frac{\dd^2 k_g}{16\pi^3}
\frac{1}{P^+}\hat\psi_{q\to qg}^{(\sigma h_1)}(k_1+k_g; k_1, k_g)
\,
\frac{1}{P^+}\hat\psi_{qg\to q}^{(\sigma h_2)}(k_2-k_g, k_g; k_2)\nn\\
& &
\frac{\Theta(1-(x_1+x_g))\, \Theta(x_2-x_g)}{(x_1+x_g)\, (x_2-x_g)}\,
\frac{1}{\sqrt 6}\epsilon_{j_1 j_2 n_3}
\Psi_\mathrm{qqq}\left(k_1+k_g;\, k_2- k_g;\,  k_3\right)~.
\eea
(There are analogous contributions corresponding to gluon exchanges
between quarks $1,3$, and $2,3$.) We now multiply by
\bea
\langle P|\alpha'\rangle =
\frac{1}{\sqrt 6} \, \epsilon_{n_1' n_2' n_3'}
\,\,
\Psi^*_\mathrm{qqq}\left(k_1'; k_2'; k_3'\right) \,\,
(2\pi)^3 \,\delta\left(1-\sum_i x_i'\right)\,\,
\delta\left(\sum_i \vec k_i'\right)\,\,
\langle S| h_1', h_2', h_3'\rangle~.
\label{eq:<P|alpha>-color}
\eea
We can trace out the quark helicities and sum over gluon polarizations
with the help of
\bea
& &
\sum_\sigma \sum_{h_1, h_2, h_3}\, \left| \langle h_1, h_2, h_3|S\rangle
\right|^2\,\, \frac{1}{P^+}\hat\psi_{q\to qg}^{(\sigma h_1)}(k_1+k_g; k_1, k_g)
\,
\frac{1}{P^+}\hat\psi_{qg\to q}^{(\sigma h_2)}(k_2-k_g, k_g; k_2) = \nn\\
& &
- \frac{(x_1+x_g)\sqrt{1- z}}{n^2+\Delta^2}\,
\frac{x_2\sqrt{1- z'}}{n^{\prime 2}+\Delta^{\prime 2}}\,
\vec n \cdot \vec n'\,
\left[ (2-z)(2-z')+zz'\langle h_1 h_2\rangle \right]~,
\eea
with $z=x_g/(x_1+x_g)$, $z'=x_g/x_2$, $\vec n = \vec k_g-z(\vec
k_1+\vec k_g)$, $\vec n' = \vec k_g - z'\vec k_2$, $\Delta^2=z^2
m^2_\mathrm{col}$, $\Delta^{\prime 2} = z^{\prime 2}
m^2_\mathrm{col}$.  This leads us to the density matrix
\bea
\rho^{(12)}_{\alpha \alpha'} &=& - \frac{g^2}{6}\, \epsilon_{n_1' n_2' n_3'}
\sum_{a, j_1, j_2} \epsilon_{j_1 j_2 n_3}\, (t^a)_{n_1 j_1}\, (t^a)_{n_2 j_2}
\int\limits_x^1 \frac{\dd x_g}{x_g}\frac{\dd^2 k_g}{16\pi^3}\,
\frac{x_2}{x_2-x_g}\, \Theta(1-(x_1+x_g))\, \Theta(x_2-x_g) \nn\\
& &
\frac{\vec n\cdot \vec n'\, \sqrt{(1- z)\, (1- z')}}
     {(n^2+\Delta^2)\, (n^{\prime 2}+\Delta^{\prime 2})}\,
     \left[(2-z)(2-z')+zz'\, \langle h_1 h_2\rangle\right]
     \,\, \Psi^*_\mathrm{qqq}(k_1'; k_2'; k_3')\,
     \Psi_\mathrm{qqq}\left(k_1+ k_g;\, k_2-k_g;\, k_3\right)
     ~.
\eea
Tracing out the quarks, we get
\bea
\tr\, \rho^{(12)} &=& \frac{g^2}{6}\, C_F N_c\,
\frac{1}{2}\int[\dd x_i] \int [\dd^2 k_i]
\int\limits_x^1 \frac{\dd x_g}{x_g}\frac{\dd^2 k_g}{16\pi^3}\,
\sqrt\frac{x_1\,x_2}{(x_1+x_g)\, (x_2-x_g)}\,
\Theta(1-(x_1+x_g))\, \Theta(x_2-x_g) \nn\\
& &
\frac{\vec n\cdot \vec n'}
     {(n^2+\Delta^2)\, (n^{\prime 2}+\Delta^{\prime 2})}\,
     \left[(2-z)(2-z')+zz'\, \langle h_1 h_2\rangle\right]
    \,\, \Psi^*_\mathrm{qqq}(k_1; k_2; k_3)
\,
    \Psi_\mathrm{qqq}\left(k_1+k_g;\,  k_2-k_g;\,  k_3\right)
     ~.
\label{eq:tr_rho12_virt}
\eea
This cancels against eq.~(\ref{eq:tr-gqqq_rho_qqqg_12'_shift}) which
is easily checked by renaming $k_1 \leftrightarrow k_2$.  The complete
UV finite virtual correction due to a gluon exchange in $|P\rangle$ or
$\langle P|$ includes $\tr\, \rho^{(13)}$, $\tr\, \rho^{(23)}$, $\tr\,
\rho^{(1'2')}$, $\tr\, \rho^{(1'3')}$, $\tr\, \rho^{(2'3')}$; this
amounts to replacing $\langle h_1 h_2\rangle \to \langle h_1 h_2 +
h_1h_3 + h_2 h_3\rangle/3 = -\frac{1}{3}$ and multiplying
eq.~(\ref{eq:tr_rho12_virt}) by a factor of 6.

\subsection{Reduced density matrix for the gluon momentum degree of freedom}
\label{sec:rho_gg'}

In this subsection we collect the expressions for the reduced
density matrix $\rho_{k_g k_g'}$ which describes the entanglement of the
momentum of the gluon with other degrees of freedom which we have traced over.
This density matrix is of block-diagonal form with the first $1\times1$
block given by the number
\bea
1 &-& 3\,\, \frac{1}{2} \int[\dd x_i]\int[\dd^2k_i] \,\,
\left|\Psi_\mathrm{qqq}(k_1; k_2; k_3)\right|^2\, \,
C_q(x_1;\, x, M_\mathrm{UV}/m_\mathrm{col}) \nn\\
&+&
4g^2\, \frac{1}{2}\int[\dd x_i] \int [\dd^2 k_i]
\int\limits_x^1 \frac{\dd x_g}{x_g}\frac{\dd^2 k_g}{16\pi^3}\,
\sqrt\frac{x_1\,x_2}{(x_1+x_g)\, (x_2-x_g)}\,
\Theta(1-(x_1+x_g))\, \Theta(x_2-x_g) \nn\\
& &
\frac{\vec n\cdot \vec n'}
     {(n^2+\Delta^2)\, (n^{\prime 2}+\Delta^{\prime 2})}\,
     \left[(2-z)(2-z')-\frac{1}{3}zz'\right]
    \,\, \Psi^*_\mathrm{qqq}(k_1;\, k_2;\, k_3)\,
\,
\Psi_\mathrm{qqq}\left(k_1+k_g;\,  k_2-k_g;\,  k_3\right)~.
\label{eq:rho_g_block1}
\eea
In the last term, $z=x_g/(x_1+x_g)$, $z'=x_g/x_2$, $\vec n = \vec
k_g-z(\vec k_1+\vec k_g)$, $\vec n' = \vec k_g - z'\vec k_2$,
$\Delta^2=z^2m^2_\mathrm{col}$, $\Delta^{\prime 2}=z^{\prime 2} m^2_\mathrm{col}$.
The expression for $C_q(x_1;\, x, M_\mathrm{UV}/m_\mathrm{col})$ is given
in eq.~(\ref{eq:Cq_x1_Lambda}).
\\

The next block is given by the sum of two contributions. The first
is the matrix
\bea
\rho_{k_g k_g'}^A &=&
8g^2\,
\frac{1}{2}\int[\dd x_i]\int[\dd^2k_i]\, \Theta(x_1-x_g)\,
\Theta(1-x_1+x_g-x_g')\, \Psi^*_\mathrm{qqq}(k_1-k_g+k_g'; k_2; k_3)
\, \Psi_\mathrm{qqq}(k_1; k_2; k_3)
\nn\\
& &~~~
\vec n \cdot \vec n' \, \frac{\sqrt{(1-z)\, (1-z')}}
     {(n^2+\Delta^2)\, (n'^2 + \Delta^{\prime 2})}
     \, (2 - z - z'+zz')      ~.
\eea
where $z=x_g/x_1$, $z'=x_g'/(x_1-x_g+x_g')$, $\vec n = \vec k_g - z\vec k_1$,
$\vec n' = \vec k_g' - z'(\vec k_1-\vec k_g+\vec k_g')$, $\Delta^2=z^2
m^2_\mathrm{col}$, $\Delta^{\prime 2} = z^{\prime 2}
m^2_\mathrm{col}$. Along the diagonal of this block one adds
\bea
&-&
8g^2\,
\frac{1}{2}\int[\dd x_i]\int[\dd^2k_i]\,
\left|\Psi_\mathrm{qqq}(k_1; k_2; k_3)\right|^2\,
     \left[1+\left(1-z\right)^2\right]\,
     \frac{1-z}{k_g^2+\Lambda^2}~,
\eea
with $\Lambda = z M_\mathrm{UV}$.
To perform the trace over this contribution one sets $k_g'=k_g$, which
implies $z'=z$ and $\vec n' = \vec n$, and integrates with the
measure~(\ref{eq:g-measure_x1}) which includes a Jacobian
$x_1/(x_1-x_g) = 1/(1-z)$. This cancels the second term in
eq.~(\ref{eq:rho_g_block1}).

The second contribution is
\bea
\rho_{k_g k_g'}^B &=& -8g^2\,
\frac{1}{2}\int[\dd x_i]\int[\dd^2k_i] \,\,
\vec n \cdot \vec n' \, \frac{\sqrt{(1-z)\, (1- z')}}
     {(n^2+\Delta^2)\, (n'^2 + \Delta^{\prime 2})}
\, (2-z-z'+\frac{1}{3}zz')\nn\\
& &
\Theta(x_1-x_g)\,\Theta(1-(x_2+x_g'))\, 
\Psi^*_\mathrm{qqq}(k_1-k_g; k_2+k_g'; k_3)\,
\Psi_\mathrm{qqq}(k_1; k_2; k_3)\,
~,
\eea
where now $z=x_g/x_1$, $z'=x_g'/(x_2+x_g')$,
$\vec n = \vec k_g - z\vec k_1$, $\vec n' = \vec k_g' -
z'(\vec k_2+\vec k_g')$, $\Delta^2=z^2 m^2_\mathrm{col}$,
$\Delta^{\prime 2} = z^{\prime 2} m^2_\mathrm{col}$. To trace this
matrix one again sets $k_g'=k_g$ and integrates with the
measure~(\ref{eq:g-measure_x1}). This cancels the last term in
eq.~(\ref{eq:rho_g_block1}).

\subsubsection{Reduced density matrix for the $x_g$ degree of freedom}
\label{sec:rho_gg'_x}

We can trace the expressions from the previous section over the gluon
transverse momentum to obtain the density matrix $\rho_{x_g x_g'}$ for
the last remaining degree of freedom corresponding to the light-cone
momentum fraction of the gluon. To render the result in as simple a
form as possible we will restrict to $x_g, x_g'$ much less than the
typical quark momentum fraction $\langle x_q\rangle$. Accordingly,
when integrating over $x_g$ we assume that the cutoff $x$ for the soft
singularity is much less than $\langle x_q\rangle$.\footnote{However,
  we also assume that $\alpha_s\log\frac{\langle x_q\rangle}{x} \ll 1$
  so that a resummation of the density matrix to all orders in this
  parameter (see ref.~\cite{Armesto:2019mna}) is not required.}
\\

The first $1\times1$ block of $\rho_{x_g x_g'}$ is then given by
\bea
1 &-& \frac{g^2}{\pi^2}\, \log\frac{\langle x_q\rangle}{x}\,
\log\frac{M^2_\mathrm{UV}}{m^2_\mathrm{col}}
+ \int\limits_x^{\langle x_q\rangle}\frac{\dd x_g}{x_g}\, F(x_g^2 m^2_\mathrm{col})~,
\label{eq:rho_g-x_block1}
\eea
with
\bea
F(x_g^2 m^2_\mathrm{col}) = 16\, g^2\, 
\frac{1}{2}\int[\dd x_i] \int [\dd^2 k_i]
\int\frac{\dd^2 k_g}{16\pi^3}\,
\frac{1}{k_g^2+\frac{x_g^2}{x_1^2} m^2_\mathrm{col}}\,
\, \Psi^*_\mathrm{qqq}(k_1;\, k_2;\, k_3)\,
\,
\Psi_\mathrm{qqq}\left(x_1,\vec k_1+\vec k_g;\,  x_2,\vec k_2-\vec k_g;\,
x_3, \vec k_3\right)~.
\label{eq:DMx_auxF}
\eea
\\

The second block is given by
\bea  \label{eq:rho-xg-xg'_2nd-block}
\rho_{x_g x_g'} &=&
\frac{g^2}{\pi^2}\, \log\frac{M^2_\mathrm{UV}}{m^2_\mathrm{col}}
- F(\mathrm{max}(x_g^2,x_g^{\prime 2})\cdot m^2_\mathrm{col})
~.
\eea
Note that taking the trace involves an integration over $\dd
x_g/x_g$. Hence, for proper normalization of the eigenvalues the
r.h.s.\ of eq.~(\ref{eq:rho-xg-xg'_2nd-block}) should be multiplied by
$\dd x_g/\sqrt{x_g\, x_g'}$ in order to transform the trace operation
to a sum over $x_g$-bins; compare to eq.~(\ref{eq:rescaled-rho_qqq}).
\\

For illustration we proceed to determine the spectrum of the above
density matrix numerically. We again use the ``harmonic oscillator''
three-quark input wave function from
ref.~\cite{Schlumpf:1992vq,Brodsky:1994fz} and set the remaining
parameters as follows: a small coupling constant
$\alpha_s=g^2/4\pi=0.1$ and a fairly large collinear regulator
$m_\mathrm{col}=1$~GeV so that the perturbative calculation should
apply, $\langle x_q\rangle = 0.3$, $\log M^2_\mathrm{UV}/
m^2_\mathrm{col}=4$, and the soft cutoff $x=0.1$.  Even for such
fairly large cutoff on the gluon light-cone momentum we obtain a low
purity of $\tr\, \rho^2 = 0.52$: two eigenvalues of the density matrix
are close to 0.5 while the others are close to 0.  This purity is
substantially lower than the purity of the reduced density matrices
for the three quark Fock state (c.f.\ table~\ref{tab:purity-wf}).

We emphasize again that the density matrix written in
eqs.~(\ref{eq:rho_g-x_block1} --
\ref{eq:rho-xg-xg'_2nd-block}) is {\em approximate}. As such,
even though it is symmetric and its trace is equal to 1 it may violate
the positivity requirement on the eigenvalues. For the set of
parameters mentioned above we find numerically that the absolute value
of the most negative eigenvalue is 50 times smaller than the smallest
positive eigenvalue.  For greater coupling $\alpha_s$ or a
substantially smaller cutoff $x$, however, the magnitude of the most
negative eigenvalue increases and so the above reduced density matrix
becomes unphysical.

\section{Summary} \label{sec:summary}

In this paper we have analyzed entanglement of degrees of freedom in
the light-cone wave function of the proton at intermediate
parton momentum fractions.  In sec.~\ref{sec:qqqDM-LO} we focused on
the three quark Fock state which should dominate for large $x$. When
one traces the pure density matrix for the anti-symmetric
$\epsilon_{i_1\cdots i_{N_c}} |i_1,\cdots, i_{N_c}\rangle$ color state
over all but one color degree of freedom then the spectrum of
eigenvalues of the resulting reduced density matrix
$\rho_{ij}=\frac{1}{N_c}\delta_{ij}$ is degenerate, and the
von~Neumann entropy is $S_\mathrm{vN}=\log N_c$, indicating maximal
entanglement of color.

On the other hand, in the limit of many colors, the spatial proton
wave function should factorize into a product of $N_c$ one-body quark
wave functions~\cite{Witten:1979kh} where spatial degrees of freedom
belonging to different quarks would not be entangled.

For $N_c=3$, we used a model three-quark wave function from the
literature~\cite{Schlumpf:1992vq,Brodsky:1994fz} to find weak
entanglement of spatial degrees of freedom (longitudinal or transverse
quark momenta); the reduced density matrices exhibit purities of
95\% or greater. These model wave function involve as the only
dimensionless physical parameter that the density matrix may depend
on, the product of constituent quark mass and proton radius, or
alternatively the mass of the proton times its radius.

However, to check whether, indeed, the known large-$x$ structure of
the proton requires weak entanglement of spatial degrees of freedom, it
would be interesting to repeat the analysis with three-quark wave
functions which actually solve a light-front Hamiltonian with
interactions~\cite{Xu:2021wwj}. Also, one could check entanglement in
light-front wave functions obtained via ``Large Momentum Effective
Theory'' from lattice
QCD~\cite{Ji:2020ect,Ji:2021znw,Liu-Zhao-Schafer-SNOWMASS21}.  \\

In sec.~\ref{sec:DM_O-g2} we included the $|qqqg\rangle$ Fock state
via light-cone perturbation theory. Tracing over quark degrees of
freedom and gluon helicity and color we obtained the reduced density
matrix for the gluon momentum degree of freedom in
sec.~\ref{sec:rho_gg'}.  In $D=4$ space-time dimensions the trace of
that density matrix receives UV divergent contributions due to the
integration over the gluon transverse momentum. Upon regularization,
the contributions from ``real emissions'' and ``virtual corrections''
cancel. However, even though the sum of eigenvalues does not depend on
the UV regulator, nor on the collinear regulator or the soft cutoff,
their spectrum does (and therefore so does the purity and the
von~Neumann entropy). In sec.~\ref{sec:rho_gg'_x} we further trace
over the gluon transverse momentum to write the reduced density matrix
for the remaining gluon light-cone momentum fraction degree of freedom
in a particularly simple form by employing a small-$x_g$ ($\ll \langle
x_q\rangle$) approximation. We obtain numerically that even for rather
weak coupling, $x_g$ appears to be more strongly entangled with the
traced-out ``environment'' than quark momentum fractions in the
three-quark Fock state. This is, at least qualitatively consistent
with the suggestion that entanglement grows stronger with decreasing
$x$~\cite{Kharzeev:2021nzh,Kharzeev:2017qzs,Kovner:2015hga,Armesto:2019mna,Dvali:2021ooc}.

\section*{Acknowledgements}

We thank Alex Kovner, Vladimir Skokov, and Raju Venugopalan for
useful discussions.
We also acknowledge support by the DOE Office of Nuclear Physics through
Grant DE-SC0002307, and The City University of New York for PSC-CUNY
Research grant 64025-00~52.  The figures have been prepared with
Jaxodraw~\cite{Binosi:2008ig}.

\bibliography{refs}

\end{document}